\documentclass[preprint,11pt]{JHEP3} 

\JHEPspecialurl{http://jhep.sissa.it/JOURNAL/JHEP3.tar.gz}
\usepackage{graphicx}
\newcommand\fverb{\setbox\fverbbox=\hbox\bgroup\verb}
\newcommand\fverbdo{\egroup\medskip\noindent%
			\fbox{\unhbox\fverbbox}\ }
\newcommand\fverbit{\egroup\item[\fbox{\unhbox\fverbbox}]}
\newbox\fverbbox

\usepackage{float}
\usepackage{amssymb,latexsym,amsmath}
\usepackage{axodraw}
\usepackage{graphicx}
\usepackage{epsfig}

\newcommand{\be}{\begin{equation}}
\newcommand{\ee}{\end{equation}}
\newcommand{\ba}{\begin{eqnarray}}
\newcommand{\ea}{\end{eqnarray}}

\def\vec#1{{\mbox{\boldmath$#1$}}}

\newcommand{\ep}{\epsilon}

\title{Second order QCD corrections to inclusive 
semileptonic 
$b \to X_c l \bar \nu_l$ decays with massless and  massive lepton}

\author{Sandip Biswas\thanks{On leave of absence from the Department of Physics and Astronomy, University of Hawaii at Manoa}\; and Kirill Melnikov \\
Department of Physics and Astronomy,
Johns Hopkins University,
Baltimore, MD 21218, USA
}

\received{} 		
\accepted{}		

\abstract{
We extend previous computations of the second order QCD corrections to 
semileptonic $b \to c $ inclusive transitions, to the case where the charged 
lepton in the final state is  massive. This allows accurate description 
of $b \to c \tau \bar \nu_\tau$ decays. We review 
techniques used in the computation of ${\cal O}(\alpha_s^2)$ 
corrections to inclusive semileptonic $b \to c$ transitions 
and present  extensive numerical studies of  ${\cal O}(\alpha_s^2)$ 
QCD corrections to $b \to c l \bar \nu_l$ decays, for $l =e, \tau$.
}

\begin{document}

\section{Introduction}

Inclusive semileptonic decays of $B$-mesons into charmed final states 
$B \to X_c l \bar \nu_l$ 
are benchmark processes in $B$-physics.  These processes were 
studied extensively at $B$-factories, LEP and 
the Tevatron \cite{exp_babar,exp_belle,exp_del,exp_cleo,exp_cdf,babar2009}.   
When lepton in the final state is electron or muon, most of the 
experimental data come from BABAR and BELLE, while measurements 
of inclusive rate for $b \to X_c \tau \bar \nu_\tau$ transition are due 
to ALEPH and OPAL \cite{expaleph,expopal}. 
At $B$-factories  only preliminary results 
on exclusive decay $B \to D  \tau \bar \nu_\tau$ were  reported
\cite{refbfacttau}.

Theoretically, inclusive semileptonic decays of $B$-mesons  
are well-understood,
thanks to the Operator Product Expansion (OPE) in inverse powers of the 
$b$-quark mass \cite{ope}. When results of experimental measurements 
for $B \to X_c l \bar \nu_{l}$ are compared 
with theoretical predictions, one is able to determine with high precision
the bottom and 
charm quark masses, the CKM matrix element $|V_{cb}|$ and the 
non-perturbative parameters of the heavy quark expansion 
\cite{fit1,fit1a,fit2}. 
Measurements of $\tau$ lepton branching fractions by ALEPH and OPAL 
are used to constrain  
possible contributions of a charged Higgs boson to
semileptonic $B$-decays.

An important issue in physics of semileptonic $B$-decays, 
intimately related to a very  
high precision achieved in experimental measurements, and the intention 
to utilize this precision fully, is the necessity to understand and control 
perturbative QCD corrections in  heavy quark decays.  
 There are two aspects of this problem. First, one  needs to understand  the
general structure of these corrections, to avoid  large perturbative 
effects. A by-now standard example of this type is the recognition 
of a special role that short-distance low-scale quark masses play 
in avoiding large perturbative effects in quark decays \cite{pt_npt}.
Second, it is very important to provide 
explicit computations of QCD radiative corrections to quantities of direct 
relevance for experimental analysis.   Large number of experimental results 
comes in the form of  moments of lepton energy, hadron energy and 
hadron invariant mass with variable cut on the lepton energy. These are 
sufficiently complicated observables to make  analytic computations, 
in particular in higher orders,  impractical. On the contrary,
if QCD effects  are computed numerically for  fully differential decay 
rates, any restriction on final state particles can be imposed.

It is interesting to point out in this regard that, while one-loop 
corrections to the decay rate and a number of basic differential distributions 
for $b \to X_c l \bar \nu_l$ decays were computed long ago 
\cite{1lrate, diffdistr}, 
fully differential decay   
rate through ${\cal O}(\alpha_s)$ was obtained 
only in 2004  \cite{trott, kolya1, kolya2}.
Early estimates of two-loop QCD corrections to the total 
rate for $b \to X_c e \bar \nu_e$ decays were given
in Refs.~\cite{mczi,mczm,mcz0}. 
Fully differential decay rate through ${\cal O}(\alpha_s^2)$
for $b \to X_c e \bar \nu_e$
was computed very recently in Ref.~\cite{melnikov}.  
That calculation is based on the techniques 
developed in Refs.~\cite{method} for the computation of 
next-to-next-to-leading order (NNLO) QCD corrections to a number 
of processes in hadron collider physics. Those techniques were 
first applied to describe weak decays of charged 
fermions  in Ref.~\cite{method1},  
where two-loop QED corrections 
to the electron energy spectrum in muon decay were studied.
We note that analytic results for semileptonic decay rate 
$b \to X_c l \bar \nu_l$ and moments of the lepton energy without 
lepton energy cut were 
computed through ${\cal O}(\alpha_s^2)$ 
in the form of $m_c/m_b$ expansion in Refs.~\cite{Pak:2008qt,Pak:2008cp}.
Analytic results reported in those references are 
very handy -- especially, when 
compared to numerical calculations reported in Ref.~\cite{melnikov} and in  
the current paper --
but it is not clear {\it a'priori} if they 
can be  used for moments with lepton energy cuts, because 
the precision achieved in experimental measurements is rather high.
Finally, we note that efforts are underway to compute ${\cal O}(\alpha_s)$ corrections 
to Wilson coefficients of 
leading non-perturbative operators, that contribute to moments in 
$b \to X_c l \bar \nu_l$ transitions \cite{tb}.

There are two goals that we pursue in this paper. First, we 
extend  the calculation 
reported in Ref.~\cite{melnikov} by considering  {\it massive} 
lepton in the final state.  Once this is accomplished, we have 
a fully differential description  of $b \to X_c l \bar \nu_l$ transition, 
where $l$ can be electron, muon  or $\tau$. 
Second, we present a much more detailed 
study of ${\cal O}(\alpha_s^2)$ corrections in semileptonic $B$-decays,
than what was published in Ref.~\cite{melnikov}.

We point out that since our computations are numerical, 
the presence of a massive lepton in the final state is 
a relatively minor complication,  so that the extension of  the calculation 
reported in \cite{melnikov} to   $b \to X_c \tau \bar \nu_\tau$ 
is straightforward.  This feature is particular 
to numeric computations  since for analytic 
calculations any additional massive particle in the final state 
is a serious complication.

We use our results for second order QCD corrections to  
$b \to X_c \tau \bar \nu_\tau$ transition 
to show that perturbative QCD corrections to the ratio 
$B(b \to X_c \tau \bar \nu_\tau)/B(b \to X_c l \bar \nu_l)$ are very small. 
This ratio was measured at LEP by the ALEPH and OPAL collaborations
with  decent precision. Non-perturbative corrections
 to this ratio are computed  
within the framework of the operator product expansion in the inverse mass 
of the $b$-quark \cite{koyrakh,neubert,Balk:1993sz} and are also rather modest. Hence, 
the ratio of the branching fractions provides a simple constraint 
on the values of bottom and charm  quark masses.  More generally, 
it appears that the ratio of two semileptonic branchings is an 
observable that can be predicted very accurately in QCD so that 
a good measurement of $ b \to X_c \tau \bar \nu_\tau$ decay branching fraction 
at future $B$-factories can be very interesting.

We also provide a detailed study of the second order QCD corrections 
to  $b \to X_c l \bar \nu_l$ for massless lepton. We compute  
the so-called non-BLM corrections to a large number of moments of 
different kinematic variables, for various charm and bottom quark masses 
and for different values of the lepton energy cut. 
These results can be used to create interpolating functions 
for ${\cal O}(\alpha_s^2)$  non-BLM corrections, for the use in fits to 
moments in semileptonic $B$ decays.

The remainder of the paper is organized as follows. 
In Section~\ref{sect:2}, 
we describe technical details of the computation. We begin by explaining the 
phase-space parametrization that we employ and then discuss details pertinent 
to our computation of two-loop virtual, real-virtual and double-real emission 
corrections. In Section~\ref{sect:3} we present the results of the calculation.
We tabulate large number of non-BLM corrections to various moments  for 
$b \to X_c e \bar \nu_e$ decay and discuss their potential impact 
on the results of global fits of semileptonic $B$-decays. 
Then, we present ${\cal O}(\alpha_s^2)$  QCD corrections 
to the decay rate $b \to X_c \tau \bar \nu_\tau$.
We conclude in Section~\ref{sect:4}. 

\section{Technical details} 
\label{sect:2}

In this Section we discuss technical aspects of the computation of 
the QCD radiative corrections to $b \to X_c l \bar \nu_l$, where lepton 
can be  electron, muon  or tau.  We would like to 
develop fully numerical 
approach to the computation of  these radiative corrections, since this is the 
only known way to achieve flexibility, required for the  
description of $b$-decays. Let us stress that 
such numerical computations would have been  
very straightforward if not for divergences that 
occur both in virtual loops and in real emission processes, when radiated 
gluons or massless quarks become soft or collinear to other particles. 

The presence of singularities makes it necessary to develop techniques 
to extract them, before proceeding to the  numerical computations.
This is accomplished with the help of the sector decomposition
\cite{sectdec}. Sector decomposition  
can be applied to integrals with complicated polynomials 
in denominators, to find   changes of variables 
that factorize all singularities of the integrand.  Given 
sufficiently complicated polynomials, such variable transformations 
can not be established globally and one needs to split the integral 
into many ``sectors'', where such variable transformations can be accomplished.
It should be stressed that the sector decomposition technique is algorithmic 
and, hence, can be easily programmed -- one does not need to examine 
all the integrals that appear in the problem to find suitable changes 
of variables.

Calculation of any physical quantity requires that  squares of matrix elements 
are integrated over  phase-space, allowed for final state particles. 
It is important,  to choose the phase-space parametrization 
which is sufficiently simple to avoid proliferation of terms in the process  
of sector decomposition. We therefore start with the detailed 
discussion of  how  
the phase-space can be parametrized.  After parametrization of 
the phase-space is fixed, we  explain how 
various parts of the NNLO computation are performed.

\subsection{Phase-space parametrization}

We need to consider phase-space parametrization for the 
following processes: {\it i}) Born  $b \to cl \bar \nu_l$; 
{\it ii}) single-gluon emission $b \to cl \bar \nu_l g$;  
{\it iii)} double-gluon emission $b \to c l \bar \nu_l g g$. In the latter 
case, the parametrization is also  valid  for quark 
emission processes $b \to c l \bar \nu_l q \bar q$, for massless quarks. 
We do not consider final states with three charm quarks 
since this contribution is  
suppressed, for realistic  bottom and charm quark 
masses.

\subsubsection{Born process: $b \to c l \bar \nu_l$}
We begin with the phase-space parametrization for Born process
$b \longrightarrow cl{\bar \nu_l}$. We label momenta of particles and 
their flavors in the same way  so that, for example, $b$ refers to $p_b$, 
the momentum of the $b$-quark, where appropriate. 
The differential phase-space for final 
state particles reads
\be
{\rm d}{\rm Lips}_{\rm LO}
=\lbrack{dc}\rbrack\lbrack{dl}\rbrack\lbrack{d\nu}\rbrack
{\left({2\pi}\right)}^{d}{\delta}^{(d)}\left(b-c-l-\bar \nu_l\right),
\ee
where the integration is performed in   $d$-dimensional space,
with  $d=4-2\epsilon$. The integration measure is  defined as
\be
\lbrack{dp}\rbrack= \frac{d^{d-1}p}{{\left(2\pi\right)}^{d-1}2{p_0}}.
\ee
The phase-space decomposition suitable for a three-body decay
 is carried out by assuming 
a sequence of two two-body decays. First,  
the $b$ quark decays into an off-shell $W$-boson  and 
the charm quark, then the 
virtual $W$-boson decays  into the lepton $l$ and the neutrino.
We write 
\be
 {\rm d}{\rm Lips}_{\rm LO}
= {\frac{d{W^2}}{2\pi}}\;
{\lbrack{dc}\rbrack\lbrack{dW}\rbrack{\left(2\pi\right)}^d\delta^{(d)}\left({b-c-W}\right)}
\;{\lbrack{dl}\rbrack\lbrack{d\bar \nu_l}\rbrack{\left(2\pi\right)}^d\delta^{(d)}\left({W-l-\bar \nu_l}\right)}.
\label{eq1}
\ee

We now parametrize all entries in Eq.(\ref{eq1}) in such a way that 
the integration region is a  unit hypercube. We begin with the leptonic 
phase-space.  We note that the leptonic phase-space is universal 
for all processes involved in the NNLO computation. Hence, 
we can choose to calculate it in four dimensions and neglect all its  
dependence on $\ep$. We therefore write 
\be
{\rm d Lips}_{W\rightarrow l+ \bar \nu_l}
= \frac{W^2-m_{l}^2}{8 \pi W^2} \; {\rm d}\lambda_{7} {\rm d}\lambda_{8},\;\;
0 \le  \lambda_{7,8} \le 1.
\ee
The physical meaning of the two parameters $\lambda_{7,8}$ is as follows -- 
they describe the polar angle $ \cos \theta_l = -1 + 2\lambda_7  $ 
and the azimuthal angle $\phi_l = 2\pi \lambda_8$, that parametrize 
lepton momentum in the
rest frame of the $W$-boson,  relative to the direction 
of the $W$ momentum in the rest frame of the decaying $b$-quark. 

The remaining phase-space that describes $b \to c +W$ transition 
can be written as 
\be
\frac{{\rm d}W^2}{2\pi}\; {\rm d Lips}_{b \to W+c} 
= \frac{\left ( \left(m_b-m_c\right)^2 -m_l^2 \right ) p_c^{d-3}}{4m_b}
\;\frac{{\rm d} \Omega_{c}}{\left(2\pi\right)^{d-1}}\;\;
{\rm d}x_1,
\ee
where ${\rm d} \Omega_{c}$ is the ($d-1$)-dimensional  solid 
angle that describes direction of the charm quark momentum $p_c$ 
 in the $b$-quark  rest frame.  The remaining kinematic 
variables  in the 
$b$-quark rest frame are written as 
\ba
&& E_c = m_c + \left ( E_c^{\rm max} - m_c \right)(1-x_1),\;\;\;
E_c^{\rm max} = \frac{m_b}{2} \left ( 1 + \frac{m_c^2-m_l^2}{m_b^2} \right ),
\nonumber \\
&& E_W = m_b - E_c,\;\;\;P_W = \sqrt{E_W^2 - W^2},\;\;\;
W^2 = m_b^2 + m_c^2 - 2m_b E_c.
\ea
It is straightforward to write the energy of the lepton  
and the angle between the lepton and the $W$ in the $b$-quark rest frame
\be
 E_{l}=\frac{E_W}{2} \left ( 1 + \frac{m_l^2}{W^2} \right ) 
- \frac{P_W}{2} \left ( 1- \frac{m_l^2}{W^2} \right ) 
 \left ( 1- 2\lambda_7 \right ),\;\;\;
\cos \theta_{lW} = \frac{2E_W E_l - W^2 - m_l^2}{2P_W p_l}.
\label{eq:lep}
\ee
In Eq.(\ref{eq:lep}), 
$p_l =  \sqrt{E_l^2 - m_l^2}$ denotes   the lepton three-momentum.

The above formulas provide sufficient information to 
compute scalar products of the four-momenta of all particles 
that appear  in the leading order calculation. 
Indeed, scalar products that involve the decaying 
$b$ quark are straightforwardly expressed in terms of particle energies 
in the $b$-quark rest frame.
The relative angle between the direction of the charged lepton 
and the direction  of the charm quark is easily related 
to $\theta_{lW}$. Neutrino 
four-momentum is given by $\bar \nu_l = W - l$ which implies that 
neutrino momentum does not lead to independent scalar products.
The phase-space 
 parametrization described in this subsection, 
is employed in all parts of the calculation where 
the three-particle phase-space enters. In NLO and NNLO computations, 
this occurs  when contributions to the decay rate due to one- and 
two-loop virtual corrections, respectively, 
are calculated.  

\subsubsection{Single gluon emission process: $b \to c l \bar \nu_l g$}
Next, we discuss the phase-space parametrization for the process
$b \to c l \bar \nu_l g$, with four particles in the final state.  When 
the energy of the emitted gluon becomes small, the corresponding 
matrix element diverges. Good parametrization of the four-particle 
phase-space  should factor out the dependence on the gluon energy, so that 
extraction of infrared divergences occurs easily. We write 
\be
{\rm d}{\rm Lips}_{\rm NLO} = 
\left [dc\right]\left[dg\right]\left[dl\right]\left[d\bar\nu_l\right]
\left(2\pi\right)^d \delta^{(d)}\left(b-c-l-\bar \nu_l-g\right), 
\ee
and decompose it into quark and lepton phase-spaces, by introducing 
the four-momentum of the $W$ boson. This leads to
\be
{\rm d}{\rm Lips}_{\rm NLO} 
=\frac{dW^2}{2\pi} {\rm d}{\rm Lips}_{b \to c + g +W}
{\rm d}{\rm Lips}_{W \to l + \bar \nu_l}.
\ee

Parametrization of the lepton phase-space is the same as for the Born 
process, described earlier.
The quark phase-space is different. To arrive at a suitable parametrization,
it is convenient to integrate over the three-momentum of the $W$-boson
and then over the energy of the gluon, 
to remove all delta-functions. We introduce three variables 
$0 < x_{1,2,3}< 1$ and write
\ba
&& E_c = m_c + \left ( E_c^{\rm max} - m_c \right)x_1,\;\;\;
E_g 
= \frac{\left[\left(m_b-m_c\right)^2-m_l^2\right]\left(1-x_1\right)x_2}
{2 (m_b - E_c + p_c \cos \theta_{cg})}, \\
&& 
W^2=m_l^2+\left[\left(m_b-m_c\right)^2-m_l^2\right]
\left(1-x_1\right)\left(1-x_2\right),\;\;\; \cos{\theta_{cg}}= -1 +2x_3.
\nonumber
\ea
All other angles can be derived. For example, the angle 
between the gluon and the $W$ in the $b$-quark rest frame reads 
\be
\cos \theta_{gW} = \frac{-E_g - p_c \cos \theta_{cg}}{P_W}.
\ee
The only other angle that we need in order to fix all independent 
scalar products is the angle between 
the lepton and the gluon, in the $b$-quark rest frame. We obtain 
\be
\cos \theta_{lg} = \cos \theta_{gW} \cos \theta_{lW} + 
\sin \theta_{gW} \sin \theta_{lW} \cos \phi_l,
\ee
where $\cos \theta_{lW}$ is given in Eq.(\ref{eq:lep}) and 
$\phi_l = 2\pi \lambda_8$.  Finally, we obtain the 
following parametrization of the NLO phase space
\ba
\frac{ {\rm d}{\rm Lips}_{\rm NLO}}{
\prod \limits_{i=1}^{3} {\rm d} x_i
{\rm d}\lambda_7 {\rm d}\lambda_8 }  &&= 
\frac{\Omega^{2} \Gamma(2-2\epsilon)}
{2^{8-2\ep} \pi \left(2\pi\right)^{2\left(d-1\right)}
\Gamma^2(1-\epsilon)}\; 
\frac{ 
p_c^{1-2\ep} \left(\left(m_b-m_c\right)^2
-m_l^2\right)^{3-2\ep}}{m_b \left( m_b - E_c + p_c \cos \theta_{cg}  
\right)^{2-2\ep}} 
\nonumber \\
&& 
\times
\left ( 1 - \frac{m_l^2}{W^2} \right ) \;
\left(1-x_1\right)^{2-2\epsilon}x_2^{1-2\epsilon}
 x_3^{-\ep} \left(1-x_3 \right)^{-\ep}.
\ea
We use $\displaystyle {\rm d}{\rm Lips}_{\rm NLO}$ 
in  NLO calculations,  
as well as for dealing with real-virtual corrections 
in NNLO calculations.

\subsubsection{Double gluon emission process:  $ b \to c l \bar \nu_l gg$}
Finally, we discuss the parametrization of the five particle phase-space 
that  is needed for the description  of 
the double real-emission processes,  such as 
$b \to c l \bar \nu_l g_1 g_2$ or $b \to c l \bar \nu_l q \bar q$.  
We introduce six variables $x_{i=1..6}$, 
that satisfy $0 < x_i < 1$ and use $x_{1,2,3}$ to parametrize  
energies of the charm quark and of the gluons,  and the $W$ invariant mass
\ba
&& E_c = m_c + \left ( E_c^{\rm max} - m_c \right)(1-x_1),\;
W^2 = m_l^2 + ( (m_b - m_c)^2 - m_l^2) x_1 (1-x_2),  
\\ 
&& E_{g_1} = \frac{ ( \left ( m_b - m_c \right )^2 -m_l^2)x_1 x_2 x_3}
{2(m_b - E_c + p_c \cos \theta_{1c}) },\;\;\;
E_{g_2} = \frac{\left ( (m_b - m_c)^2 - m_l^2 \right ) 
x_1 x_2(1-x_3)}{2(m_b -E_c + p_c \cos \theta_{2c}) - E_{g_1} (1 
- \cos \theta_{12})}. \nonumber
\ea
We use $x_{4,5,6}$ to parametrize the relevant angles.   
We note that, in order to  handle collinear 
singularities related to $g^* \to gg $
splitting,  
it is useful to have a simple parametrization of the relative angle 
between the two gluons.  Therefore, we choose the $z$-axis to be 
aligned with the momentum of the gluon $g_1$; we choose the $x$-axis in such 
a way that the gluon $g_2$ is in the $z-x$ plane.
This fixes the global reference frame. Then, we  introduce the 
relative angles between two gluons, gluon $g_1$ and the charm quark 
and the azimuthal angle of  the charm quark $\phi_c$
\be
\cos \theta_{12} = 1 - 2 x_4;\;\;\; \cos \theta_{1c} = 1 - 2x_5; \;\;\;
\cos \phi_{c} =  1 - 2\sin^{2} \left ( \frac{\pi x_6}{2} \right ).
\ee
Given those angles, we can find all other angles between momenta of different 
particles. For example,  the angle between the charm quark 
 and the gluon $g_2$ reads  
\be
\cos \theta_{c2} = \sin \theta_{12} \sin \theta_{1c} \cos \phi_c 
+ \cos \theta_{12} \cos \theta_{1c}.
\ee

The angle between the $W$-boson  and any other particle is computed from 
momentum conservation $W = b - c - g_1 - g_2$. For example, the 
angle between the $W$-boson  and the gluon $g_1$ is given by 
\be
\cos \theta_{1W} = \frac{-E_{g_1} - E_{g_2} \cos \theta_{12} 
- p_c \cos \theta_{1c}}{P_W}.
\ee
The relative 
angle between the charged lepton and  any other particle 
is derived 
in a similar way. For example,  the angle between the gluon $g_1$ and the 
lepton $l$ reads 
\ba
\cos \theta_{1l} = \cos \theta_{lW} \cos \theta_{1W} + 
\sin \theta_{1W} \sin \theta_{lW} \cos (\phi_l).
\ea

Finally, we express the phase-space 
for $b \to c l \bar \nu_l g_1 g_2$ decay through the appropriate variables 
and obtain
\ba
&& \frac{{\rm d Lips}_{\rm NNLO}}{\prod \limits_{i=1}^{6}{\rm d}x_i {\rm d}\lambda_7 {\rm d}\lambda_8} =
\frac{\Omega^3 \Gamma^2(2-2\ep) \Gamma(1-2\ep)}{2^{11-4\ep} (2\pi)^{3d-3}\Gamma^4(1-\ep) \Gamma^2(1/2-\ep)}
\left ( 1 - \frac{m_l^2}{W^2} \right ) 
 \\ 
&& \times 
\frac{((m_b - m_c)^2-m_l^2)^{5-4\ep} p_c^{1-2\ep}}
{m_b 
(m_b - E_c + p_c \cos \theta_{1c})^{2-2\ep}
(m_b -E_c + p_c \cos \theta_{2c} - E_{g_1} (1 
- \cos \theta_{12}))^{2-2\ep}
}
\nonumber \\
&& \times 
x_1^{4-4\ep}x_2^{3-4\ep}x_3^{1-2\ep} 
(1-x_3)^{1-2\ep} x_4^{-\ep}(1-x_4)^{-\ep} 
x_5^{-\ep}(1-x_5)^{-\ep}x_6^{-\ep}(1-x_6)^{-\ep}.
\nonumber
\ea

\begin{figure}[t]
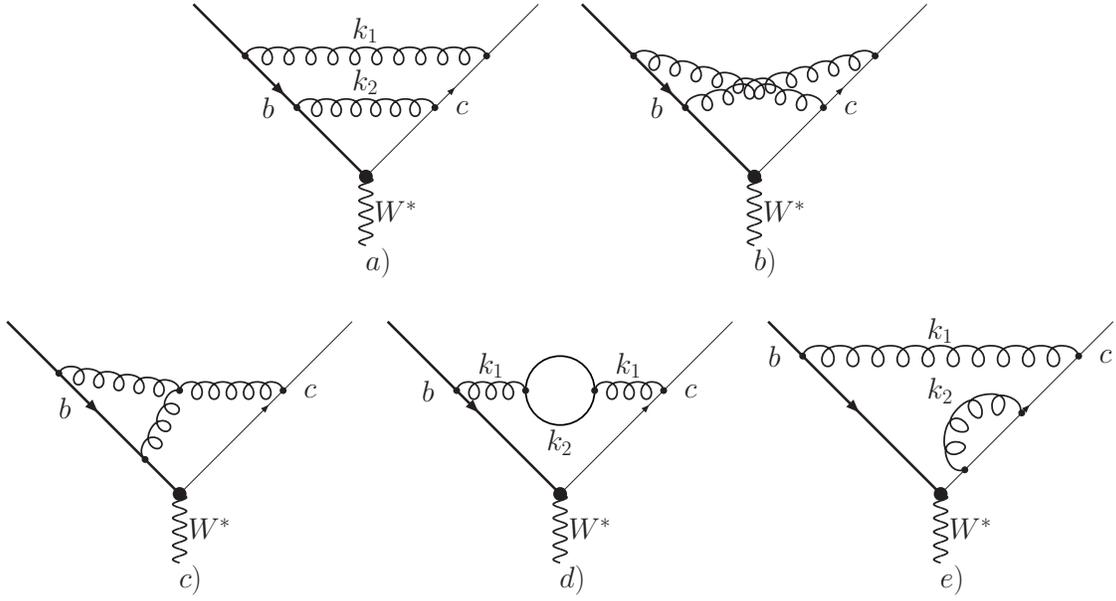

\begin{center}
\includegraphics[angle=0,scale=0.65]{bdec1_fig1.epsi}\;\;\;\;
\includegraphics[angle=0,scale=0.65]{bdec1_fig2.epsi}\\

\vspace*{0.5cm}
\includegraphics[angle=0,scale=0.65]{bdec1_fig3.epsi}\;\;\;\
\includegraphics[angle=0,scale=0.65]{bdec1_fig4.epsi}\;\;\;
\includegraphics[angle=0,scale=0.65]{bdec1_fig5.epsi}
\caption{
Sample two-loop diagrams that contribute to $b \to c +W$ transition.}
\label{fig1}
\end{center}
\end{figure}

\vspace*{0.2cm}
Having discussed parametrization of phase-spaces that we employ in 
the NNLO computation, we continue 
with the description of  technical details relevant for the  computation 
of ${\cal O}(\alpha_s^2)$ QCD corrections to  
$b \to c l \bar \nu_l$ transition.  There are three distinct 
components  that need to be addressed  -- two-loop 
virtual corrections, virtual corrections to single real emission 
process and double real emission corrections. Since these components 
require different techniques, we describe the relevant  details 
in the following subsections.

\subsection{Two-loop virtual corrections}

We begin with the discussion of how two-loop virtual corrections 
are computed. There are twelve two-loop diagrams; examples are 
shown in Fig.\ref{fig1}. These diagrams are complicated  because 
they involve several mass scales, $m_b^2, m_c^2, W^2, m_l^2$ as well 
as complicated tensor integrals, e.g. due to 
spin correlations of final state leptons 
with bottom and charm quarks. These features make
analytic computations impractical. However,  
we can compute those diagrams numerically using the method of 
sector decomposition \cite{sectdec}.  
We point out that application of sector 
decomposition is simplified in $b$-decays,
since  the two-loop diagrams 
do not develop genuine imaginary parts. 
In principle, sector 
decomposition  method was extended recently \cite{buka,buka1} to deal 
with problems where imaginary parts do appear but it is 
a welcome feature of the problem at hand, that  we do not need to deal 
with additional complications.

Hence,  the primary issue in the calculation of two-loop Feynman 
diagrams that we have to address is the efficient choice 
of  Feynman parameters,  to 
reduce the amount of sectors that are created in the process of 
sector decomposition. 
It is also important to perform  integration 
over loop momenta in such a way that computation of 
tensor integrals does not introduce kinematic singularities.
It turns out that a very simple and fairly efficient 
way to deal with tensor integrals 
is to integrate over two loop momenta sequentially.
To illustrate this procedure, we consider the planar two-loop 
diagram, shown  in Fig.1a. This diagram can be represented as a linear 
combination of tensor integrals  
\begin{equation}
I_{ij}= \int{\frac{d^{d}k_1}{\left(2\pi\right)^d}}
\int{\frac{d^{d}k_2}{\left(2\pi\right)^d}}
\frac{\{k_1\}_i\{k_2\}_j}{D_1D_2D_3D_4D_5D_6},\\
\end{equation}
where $k_1$ and $k_2$ are the loop momenta, $\{q\}_i = 
q_{\mu_1}...q_{\mu_i}$ is the rank-$i$ tensor, composed of the relevant 
loop momenta and $D_{m=1..6}$ 
are inverse Feynman propagators that appear in the planar diagram. They read
\ba
&& D_1= {k}^2_1, \qquad D_2 ={k}^2_1 + 2k_1b, \qquad  
D_3 = {k}^2_1 + 2k_1c,    \\
&& D_4={k}^2_2,\qquad  
D_5 = k_{12}^2+2k_{12} b, \qquad
D_6 = k_{12}^2+2k_{12}c,
\ea
where $k_{12} = k_1 + k_2$ is used. 
To integrate  over momentum $k_2$,  we  
introduce two Feynman parameters and write
\be
\int \frac{d^dk_2}{(2\pi)^d}
\frac{\{k_2 \}_j}{D_4D_5D_6}
= 2 \int \limits_{0}^{1} [{\rm d}x_2{\rm d}x_3] 
\int\frac{d^dK}{(2\pi)^{d}}\frac{\{K - Q\}_j}{\left(K^2-\Delta(k_1)\right)^3},
\label{eq15}
\ee
where $ [{\rm d}x_2 {\rm d}x_3] = {\rm d} x_2 {\rm d} 
x_3 \theta(1-x_2 -x_3)$ is the 
integration measure and  
\ba
&& K= k_2+ Q,\;\;\; Q = \left(k_1+b\right)x_2 + \left(k_1+c\right)x_3, \\
&& \Delta(k_1)  =  Q^2 - \left({k}^2_1 + 2k_1 b\right)x_2 
- \left({k}^2_1 +2k_1 c\right)x_3. 
\ea
Integration over the shifted loop momentum $K$ is standard and can be 
easily performed for arbitrary rank tensor.  The important  point  is 
that the higher the rank of the tensor, the smaller the power of the 
function $1/\Delta$ is in the resultant integral. Since
numerical integration is mostly problematic because 
of  infrared divergences, we should be looking  at 
the most infrared-singular 
integral,  which is provided by  the $K$-less term 
 in the numerator of Eq.(\ref{eq15}). For such integrals 
we find 
\be
2 \int \limits_{0}^{1} [{\rm d}x_2{\rm d}x_3]
\int\frac{d^dK}{(2\pi)^d}\frac{\{Q\}_j}{\left(K^2-\Delta(k_1)\right)^3}
=  - \frac{i\Gamma\left(1+\ep \right)}{(4\pi)^{d/2}}
\int [{\rm d}x_2 {\rm d}x_3]
\frac{\{ Q \}_j }{\left [ \Delta(k_1) \right ]^{1+\ep}}.
\label{eq166}
\ee
Note that integrals with higher powers of $K$ in 
Eq.(\ref{eq15}),   can, after $K$-integration,  be written 
as a linear combination of the integrands in the right hand side 
 of Eq.(\ref{eq166}),
by multiplying and dividing the integrand by appropriate 
powers of $\Delta(k_1)$. The important point is that additional 
powers of $K$ in Eq.(\ref{eq15}) do not generate yet higher powers of 
$1/\Delta$ in Eq.(\ref{eq166}).

The next step requires integration over 
$k_1$; to do that, it is convenient to change variables 
$x_2 = \lambda_1 \lambda_2,\;x_3 = \lambda_1 (1-\lambda_2)$. 
The integral over $k_1$ becomes 
\be
I_{ij} \to I_{i_1} = \int 
\; \frac{{\rm d} \lambda_1 {\rm d} \lambda_2 }
{\lambda_1^{\ep} (1-\lambda_1)^{1+\ep}}
\int \frac{d^d k_1}{(2\pi)^d}
\frac{\{ k \}_{i_1}}{D_1 D_2 D_3 {\tilde \Delta }^{1+\ep}},
\ee
where 
\be
\tilde \Delta = k_1^2 + 2k_1 Q_2 
- \frac{\lambda_1 Q_2^2}{1-\lambda_1}, 
\ee
and 
\be
Q_2 = \lambda_2b +\left(1-\lambda_2\right)c.
\ee

Since $\tilde \Delta$ is a  polynomial in $k_1$, 
the integration over $k_1$ can be performed in, essentially, the same  way 
as what was described above in regard with $k_2$ integration.   
Integrals with strongest infrared divergences  are the ones 
without additional powers of $k_1$ in the numerator. 
The corresponding  scalar integral reads 
\be
I_{0} = \frac{\Gamma(2+2\ep)}{(4\pi)^{d}}
\int \limits_{0}^{1} \prod \limits_{i=1}^{5} {\rm d}\lambda_i 
\frac{ \lambda_1^{-\ep}(1-\lambda_1)^{1+\ep} 
\lambda_4 \lambda_5^{\ep}(1-\lambda_5)^{2}}{
F^{2+2\ep}},
\label{eq167}
\ee
where 
\be
F = (1-\lambda_1)(Q_3\lambda_4 (1-\lambda_5) + Q_2 \lambda_5)^2 
+Q_2^2 \lambda_1 \lambda_5 
\label{eq168}
\ee
and 
\be
Q_3 = \lambda_3 b + (1-\lambda_3) c.
\ee

It is clear from Eqs.(\ref{eq167},\ref{eq168}) 
that the function $1/F^{2+2\ep}$ develops overlapping 
singularities at the integration 
boundaries; 
for example $F=0$ for $\lambda_1 = 1, \lambda_5 = 0$ and for 
$\lambda_4 = 0, \lambda_5 = 0$.
To disentangle those singularities, we employ the technique of 
sector decomposition \cite{sectdec}. To this end,  we map 
all the singularities to the origin 
by splitting the integration region into two intervals 
 [0,1/2] and [1/2,1], for each $\lambda_i$, 
and then change variables $\lambda_i \to \lambda_i/2$ 
and $\lambda_i \to  1 - \lambda_i/2$ in the first and second interval, 
respectively. The sector decomposition is then applied to the integrand; 
this allows us to find a sequence of variable transformations   
that factorize all singularities. Once singularities are 
factored out,  for each tensor integral 
we get expressions of the following type 
\be
I_{ij}  = \sum \limits_{\alpha \in {\rm sect}}^{} 
\int \limits_{0}^{1} 
{\rm d} \lambda_1 {\rm d} \lambda_2 ...{\rm d} \lambda_n 
\frac{N^{\alpha}_{ij}(\lambda_1,\lambda_2...\lambda_n)}
{\prod \limits_{i=1}^{n_\alpha} \lambda_i^{1+a_i\ep} D_{ij}^{\alpha}(\lambda_1,\lambda_2,...\lambda_n)}, 
\ee
where all functions $N_{ij}^{\alpha}(\{\lambda_i \})$ 
and $D_{ij}^{\alpha}(\{\lambda_i\})$ 
are finite throughout the integration region. 
Hence, all the singularities of the integrand 
are in explicitly factorized form and 
it is easy to obtain integrable expressions 
by employing  the plus-distribution prescription
\be
\frac{1}{\lambda^{1+a\epsilon}} 
=\frac{-1}{a \epsilon}\delta\left(\lambda\right) +
\sum_{n=0} {\left[{\frac{\ln^n\left(\lambda\right)}{\lambda}}\right]}_+ 
\frac{(-a \epsilon)^n}{n !},
\label{eq:plus}
\ee
where 
\be
\int_{0}^{1} d\lambda f\left(\lambda\right)
{\left[{\frac{\ln^n\left(\lambda\right)}{\lambda}}\right]}_+ =  
\int_{0}^{1} d\lambda\frac{f\left(\lambda\right)-f\left(0\right)}{\lambda}\ln^n\left(\lambda\right).
\ee

\vspace*{0.5cm}
We are now in a position to sketch all the steps that we go through to 
carry out  a calculation of a planar two-loop diagram. 
After the planar two-loop  diagram is multiplied with 
the complex conjugate tree-level amplitude and summation over polarizations 
of  external particle is performed, we use Form \cite{form} 
to integrate over  
the loop momenta, following the procedure that we just described. 
As explained earlier, we can always write the 
result in a form similar to that of the scalar two-loop integral provided 
that we allow for a polynomial function 
of Feynman parameters in the numerator. Since those numerator 
functions are finite, we do not need to know their explicit form and 
can treat them as generic finite 
functions in the process of sector decomposition. The sector decomposition procedure is coded 
up in Maple \cite{maple}.   
After sector decomposition is completed, 
Fortran files that contain finite functions to be integrated
and all the changes of variables that  
the sector decomposition procedure found necessary to apply to the 
integrand to factor out potential singularities, 
are automatically written out.

Computation of the non-planar diagram and 
the diagram with the three-gluon vertex is very similar to what 
is described above; all that changes is the Feynman parametrization. 
However, the procedure has to be modified for  diagrams 
with vacuum polarization insertions on a gluon line 
and  for diagrams with self-energy insertions on bottom and charm quark 
lines.  
We begin with the discussion of 
the vacuum polarization diagrams
with massless particles, e.g.  gluons, quarks and  ghosts. Such diagrams 
read
\be
I_{\rm VP}  = 
\int{\frac{d^{d}k_1}{\left(2\pi\right)^d}}
\int{\frac{d^{d}k_2}{\left(2\pi\right)^d}}
\frac{N(k_1,k_2)}{{k}_2^2\left(k_1+k_2\right)^2\left({k}_1^2\right)^2
\left({k}_1^2+2k_1b\right)\left({k}_1^2+2k_1c\right)}.
\label{eq:vp}
\ee
An obvious issue here is the presence of two identical gluon 
propagators  $k_1^{-2}$. As the result, in the limit 
$k_1 \to 0$, the denominator in Eq.(\ref{eq:vp}) develops  
cubic, rather than linear, singularity. In principle, even in this case, 
it is possible to proceed 
along the lines described above for the planar diagram all the way 
through the application of the sector decomposition and factorization 
of singularities. However, the complication occurs
in the process of the extraction of singularities using plus-distributions
Eq.(\ref{eq:plus}),  
since a term that scales like  $x^{-1-n}$ for $x \to 0$,  
leads to an expansion that involves 
$n$-th {\it derivative} 
of a $\delta$-function rather than the $\delta$-function 
itself.  As it turns out, 
this complication is 
unnecessary since one can analytically integrate  over $k_2$ in  any massless 
vacuum polarization 
diagram  and observe that
\be
\int{\frac{d^{d}k_2}{\left(2\pi\right)^d}} 
\frac{N(k_1,k_2)}{{k}_2^2\left(k_1+k_2\right)^2} \sim k_1^2 \tilde N(k_1),
\ee
thanks to gauge-invariance. When dealing with massless vacuum 
polarization diagrams, we indeed integrate over $k_2$ analytically, 
cancel one of the $1/k_1^2$ propagators and then perform  numerical 
integration over $k_1$. 

Clearly, a similar problem occurs also in the case of vacuum polarization 
corrections with massive quarks. In that case, however, it is harder 
to explicitly factor out  the dependence on the loop momentum $k_1$, 
to cancel the cubic divergence at small $k_1$.  For vacuum polarizations 
with massive quarks we adopt a different strategy -- we subtract those 
vacuum polarization loops  at zero momentum transfer and use 
dispersion representation to connect the two-loop diagram with the massive 
fermion loop to a one-loop diagram with the massive gluon \cite{voloshin}. 

Non-integrable singularities appear also   
in  diagrams with 
the self-energy insertion on the massive ($b$ or $c$) lines; in this 
case they are  caused by the square of the massive 
propagator which becomes nearly on-shell for small momentum of the virtual 
gluon.
In variance with the case of the  massless vacuum polarization,  
it is not possible to perform analytic integration over the loop 
momentum of the  ``self-energy'' loop.  To get around 
this problem, we use a particular integral representation 
for the quark self-energy diagram.  We consider a  
self-energy diagram  
\be
\hat \Sigma = 
\int{\frac{d^{d}k_2}{\left(2\pi\right)^d}}
\frac{\gamma^\mu ( \hat p + \hat k_2 + m_i) \gamma_\mu}{((p+k_2)^2-m_i^2) k_2^2},
\ee 
where $m_i$ stands for  $m_{b}$ or $m_c$. We 
combine two denominators using Feynman parameters,  integrate 
over the loop momentum and  obtain 
\be
\hat \Sigma = 
\frac{i \Gamma(\ep)}{(4\pi)^{d/2}} 
\int \limits_{0}^{1}{\rm d} x x^{-\ep}(p^2-m_i^2)^{-\ep}
\frac{\gamma^\mu ( \hat p +  m_i - \hat p x) \gamma_\mu}
{(1 - p^2x/(p^2-m_i^2))^{\ep}}.
\ee
This integral can be written through hypergeometric functions. 
To this end, we introduce two Dirac structures 
\be
\hat N_1 = \gamma^\mu ( \hat p +  m_i) \gamma_\mu = (2-d)\hat p + d m,\;\;\;
\hat N_2 = \gamma^\mu \hat p \gamma_\mu = (2-d) \hat p,
\ee 
and write 
\ba
\Sigma && = \frac{i \Gamma(\ep)}{(4\pi)^{d/2}} 
(p^2-m_i^2)^{-\ep} \left [ \hat N_1\; \;  \frac{\Gamma(1-\ep)}{\Gamma(2-\ep)} 
F_{21}\left (\ep,1-\ep;2-\ep,\frac{p^2}{p^2-m_i^2} \right )
\right. \nonumber \\
&& \left. - \hat N_2 \;\; \frac{\Gamma(2-\ep)}{\Gamma(3-\ep)}
F_{21}\left (\ep,2-\ep;3-\ep,\frac{p^2}{p^2-m_i^2} \right )
\right ].
\ea
These hypergeometric functions are not suitable for the purpose of 
subsequent integration because the on-shell limit $p^2 \to m_i^2$ 
is at infinity.  To take care of that, it is useful to employ an identity 
that relates the hypergeometric functions with the argument $z$ and 
$z/(z-1)$.  If we perform this transformation and go back to the integral 
representation, both of the hypergeometric functions (HGFs) become 
\be
(p^2-m_i^2)^{-\ep}{\rm HGFs} \to 
\int \limits_{0}^{1} {\rm d} x \frac{x^{\ep - a} (1-x)^{b-2\ep}}{p^2 - m_i^2/x},
\label{eq242}
\ee
where $a,b$ are some integers. 
It is now straightforward to subtract from  and add to the integrand 
in the right hand side
of Eq.(\ref{eq242})  its value at $p^2 = m_i^2$ 
\be
\int \limits_{0}^{1} {\rm d} x \frac{x^{\ep - a} (1-x)^{b-2\ep}}{p^2 - m_i^2/x}
=
\frac{p^2 - m_i^2}{m_i^2}\; 
\int \limits_{0}^{1} {\rm d} x \frac{x^{\ep - a + 1} (1-x)^{b - 1 - 2\ep }
}{p^2 - m_i^2/x} 
- m_i^{-2} B(\ep-a + 2,b-2\ep).
\label{eq195}
\ee
The first term in the right hand side of Eq.(\ref{eq195})
can be inserted in 
a two-loop diagram since the pre-factor $p^2-m_i^2$ cancels one 
of the problematic massive fermion 
propagators. The second term is a constant; it  can be combined with 
the mass counter-term contribution to cancel quadratic singularity 
and, after that, calculated explicitly.


\subsection{Mixed real-virtual corrections}
In this subsection we discuss the calculation  of 
one-loop radiative corrections 
to single gluon real emission amplitudes. At first sight, these corrections 
may look much  simpler   than the two-loop virtual corrections 
discussed previously, since they involve only one-loop virtual diagrams 
and the final state with relatively low multiplicity. It would appear 
therefore that, technically, they fall into a category of well-established
next-to-leading (NLO) calculations \cite{pv}.  Unfortunately, this 
is not quite true since, in contrast to standard NLO computations, 
the emitted  gluon in the 
final state can become soft, invalidating applicability of NLO computational
techniques. Therefore, real-virtual corrections  require careful 
study.

\begin{figure}[t]
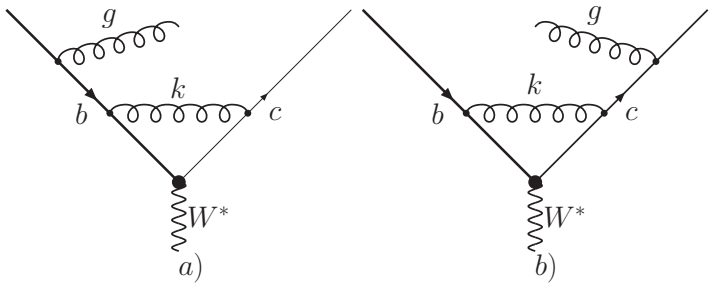

\begin{center}
\includegraphics[angle=0,scale=0.65]{bdec1_fig6.epsi}
\includegraphics[angle=0,scale=0.65]{bdec1_fig7.epsi}\\
\caption{Sample  diagrams that describe one-loop corrections 
to $b \to c +W + g$ transition.}
\label{fig2}
\end{center}
\end{figure}

There are two strategies that one can pursue to deal with the real-virtual 
corrections. One can use Passarino-Veltman tensor 
reduction technique \cite{pv} and integration-by-parts 
identities \cite{ibp} 
to reduce real-virtual corrections to one-loop scalar integrals. 
Then,  one can  attempt to extract singularities that appear when the 
energy of the gluon in the final state becomes small. While this approach 
was used in a number of calculations \cite{method,method1},
it rapidly becomes impractical with the increase in the number of particles 
in the final state.

A flexible method should be based on numerical computations and  
it seems that Feynman parametrization of one-loop virtual corrections 
and subsequent application of sector decomposition to both Feynman parameters 
{\it and} the energy of the emitted gluon is a straightforward thing 
to do. The only problem with that approach 
is that one-loop corrections to real gluon 
emission do develop imaginary parts, even when all parameters in the integrand 
are real. Technically this happens because of the singularity on the 
real integration axis which is regulated by the $+i0$ prescription.
While it is easy to implement such a prescription in analytical
computations, it is difficult to do so in a fully numerical approach.  

It turns  out that there is a simple way to avoid the issue  of the 
imaginary part in this problem. To this end, we observe that, 
for any Feynman diagram of the 
real-virtual type that contributes to $b \to c$ transition, it is possible 
to choose integration 
variables in such a way that the integration over at least one variable 
is of the form 
\be
I = \int {\rm d} \vec x \int \limits_{0}^{1} 
{\rm d} y F(\vec x, y),\;\;\;
F(\vec x , y) = \frac{y^{n_1-n_2 \epsilon}}{\left(-A(\{{\bf x}\}) 
 +B(\{{\bf x}\}) y + i 0 \right)^{n_3+ n_4 \epsilon}}.
\label{eq135}
\ee
In Eq.(\ref{eq135}) 
 $\vec x$ is a collection of other variables involved in the computation 
of the integral, $A$ and $B$ are some functions of those variables
which satisfy $B(\{ \vec x \}) > A(\{ \vec x \} )$ for all $\vec x$ 
and $n_{i=1..4}$ are integers.  We do not have a proof that such a
parametrization is possible for real-virtual corrections under all 
possible circumstances, 
but we find empirically that it exists
for $b \to c$ transitions.  

The problem with the numerical evaluation 
of  the integral in Eq.(\ref{eq135}) is that, for $n_3 > 0$, 
it becomes  
singular at $y = A(\vec x)/B(\vec x) < 1$, so that this singularity 
occurs in the middle of the integration region. Such singularity can not 
be included into the 
sector decomposition framework in a straightforward way. 
To deal with this problem, we rewrite Eq.(\ref{eq135}) 
in the following manner
\be
I = \int {\rm d} \vec x \int \limits_{0}^{\infty} {\rm d} y \;
F(\vec x, y) 
-
\int {\rm d} \vec x \int \limits_{1}^{\infty} 
{\rm d} y\; F(\vec x, y),
\label{eq136}
\ee
and observe that the first integral can be computed {\it analytically}, while 
denominator of the function $F(x,y)$ in the second term is sign-definite. 
Changing variables $y \to 1/y$ in the second term in Eq.(\ref{eq136}),
we obtain the integral that  is amenable to sector decomposition.

We now illustrate this general discussion by considering 
explicit examples.  To set the stage, we 
begin with  a simple  case, where the imaginary part problem 
does not occur. This happens  for all diagrams  where the 
gluon is emitted from the 
$b$-quark line. For our example, we consider diagram Fig.~\ref{fig2}a.
Interference of this diagram with the tree amplitude 
that describes radiative decay of the $b$-quark $b \to c + W +g$, contributes 
to ${\cal O}(\alpha_s^2)$ correction to the decay rate. We 
consider a scalar integral associated with this loop diagram
\be
I = \frac{1}{\Gamma(1+\ep)} \int \frac{d^dk}{i\pi^{d/2}}
\frac{1}
{k^2\left(k^2+2k\left(b- g\right)-2b g\right)\left(k^2+2k c\right)
}.
\ee
We introduce Feynman parameters and  
integrate over the loop momentum to obtain 
\be
I = \int 
\frac{[{\rm d}x_2{\rm d}x_3]}
{\left[\left(\left(b-g\right)x_2+cx_3\right)^2 + 2b  g x_2\right]^
{1+\epsilon}}
= 
\int \limits_{0}^{1} {\rm d} \lambda_1 {\rm d}\lambda_2
\frac{\lambda_1^{-\ep}}
{\left [ \Delta(\lambda_1,\lambda_2) \right ]^{1+\ep}},
\ee
where 
\be
\Delta(\lambda_1,\lambda_2) = 
\lambda_1 \left(\left(b-g\right)\lambda_2+c(1-\lambda_2) \right)^2 
+ 2b  g \lambda_2.
\ee
Writing $b - g = c +W$, we 
find 
\be
\Delta = \lambda_1 ( m_c^2 + 2cW \lambda_2 + W^2 \lambda_2^2) 
+ 2bg \lambda_2 > 0,
\ee
for all values   $0 < \lambda_1,\lambda_2 < 1$.
This representation for $\Delta(\lambda_1, \lambda_2)$ is instructive 
since it shows how new overlapping singularities appear when the emitted 
gluon becomes soft. Indeed, for non-vanishing gluon energy, 
$\Delta(\lambda_1,\lambda_2)$ vanishes for  $\lambda_1 =  \lambda_2 
 = 0$. On the other hand, if the 
gluon is soft $g \to 0$, $\Delta(\lambda_1,\lambda_2)$ 
vanishes for $\lambda_1 = 0$ and {\it any} value of  $\lambda_2$. To take 
care of all possible cases, we employ explicit parametrization of the 
gluon energy in the $b$-quark rest frame, as described in  the previous Section,
and  perform sector decomposition of the amplitude squared treating 
$\lambda_1, \lambda_2$ and $b g$ on equal footing. This allows us 
to extract all singularities associated with vanishing of 
{\it both}, the loop momentum 
and the momentum of the gluon in the final state.

We now turn to the description of a more difficult case which 
occurs when, in the one-loop amplitude, 
 the 
gluon is emitted from the charm quark line.   A representative diagram is 
shown in Fig.\ref{fig2}.  All the problems that appear in this  case can 
be illustrated by considering the scalar integral  as an example. We have 
\ba
&& I = \frac{1}{\Gamma(1+\ep)} 
\int \frac{d^dk}{i\pi^{d/2}}
\frac{1}{k^2\left(k^2+2kb\right)
\left((k+c+g)^2 -m_c^2\right)} \nonumber \\
&& = \int 
\frac{[{\rm d}x_2 {\rm d}x_3 ]}{\left [ \left(bx_2+\left(g+c\right)x_3\right)^2 -2gcx_3\right]
^{1+\epsilon}}.
\ea
Changing variables  $x_2=\lambda_1(1-\lambda_2)$ and 
$x_3= \lambda_1 \lambda_2 $, we obtain 
\be
I = \int {\rm d}\lambda_1 {\rm d}\lambda_2 \frac{{\lambda}_1^{-\epsilon}}
{\left ( \Delta  \lambda_1 -2 cg \lambda_2
\right )^{1+\epsilon}} 
= \int {\rm d}\lambda_1 {\rm d}\lambda_2 \frac{{\lambda}_1^{-\epsilon}}
{\Delta^{1+\ep}\left ( \lambda_1 - \xi
\right )^{1+\epsilon}},
\label{eq234}
\ee
where
\be
 \Delta = (b - W\lambda_2 )^2,\;\;\; 
\xi = \frac{2cg \lambda_2}{\Delta} < 1.
\ee
It is clear from 
Eq.(\ref{eq234}) that there is a singularity at $\lambda_1 = \xi$, i.e. 
in the middle of the  integration region, 
which can not be dealt with using 
the sector decomposition.  To transform Eq.(\ref{eq234}) 
into a form suitable for numerical integration, we  proceed 
along the lines described in the beginning of this Section. 
Consider integration  over $\lambda_1$. Allowing for a
general form of the integrand, we re-write 
\ba
&& \int \limits_{0}^{1} {\rm d} \lambda_1 
\frac{\lambda_1^{a - \ep} }{(\lambda_1 - \xi)^{b+\ep}} 
= 
\int \limits_{0}^{\infty} {\rm d} \lambda_1 
\frac{\lambda_1^{a - \ep} }{(\lambda_1 - \xi)^{b+\ep}}  
- 
\int \limits_{1}^{\infty} {\rm d} \lambda_1 
\frac{\lambda_1^{a - \ep} }{(\lambda_1 - \xi)^{b+\ep}} 
\nonumber \\
&& = 
\xi^{1+a-b-2\ep} \Gamma(1-b-\ep)
\left ( 
(-1)^{b+\ep} \frac{\Gamma(a+1-\ep)}{\Gamma(2+a-b-2\ep)}
+ \frac{\Gamma(2\ep-a+b-1)}{\Gamma(\ep-a)}
\right ) \nonumber \\
&& - 
\int \limits_{0}^{1} {\rm d} \lambda_1 
\frac{\lambda_1^{b-a + 2\ep-2} }{(1- \xi \lambda_1)^{b+\ep}} .
\label{eq235}
\ea
The remaining integrand is sign-definite, because $\xi < 1$.  
The problem of the singularity in the middle of the integration region 
has been handled by an analytic integration; the remnant of this 
problem is the imaginary part in $(-1)^{b + \ep}$.  
It is now very straightforward 
to apply the sector decomposition to the remaining integral
in Eq.(\ref{eq235}), to extract 
the infrared singularities. 

The above technique is applicable to all real-virtual diagrams. 
The first step -- finding  Feynman parametrization that is 
linear in (at least) 
one variable -- requires 
careful inspection of each diagram individually  but, once such parametrization 
is found, remaining steps are easily accomplished using algebraic manipulation 
programs. A slightly different approach is required for real-virtual 
contributions related to self-energy insertions on massive fermion lines
where quadratic singularities appear.  
We deal with those singularities using parametrization of self-energy diagrams 
described earlier in this Section in the context 
of two-loop virtual corrections.

\subsection{Double real emission corrections}

Finally, we address the computation of the double real emission 
corrections. Because we deal 
primarily  with massive quarks, 
the majority of diagrams develop only infrared singularities.
We can easily extract those using explicit parametrization of gluon 
energies in the $b$-quark rest frame, discussed earlier in this Section.
The only exception to the ``no collinear singularity'' rule comes from 
diagrams where an off-shell gluon splits into two gluons or a 
$q \bar q$ pair.  In principle, since we use the relative angle between 
two gluons (or massless quarks) as one of the primary variables 
for the phase-space parametrization, extraction of (potential) collinear 
singularity is also straightforward. However, the complication arises 
because of the necessity to deal with the extraction of quadratic singularities 
in some of the diagrams.  The problem and the solution is best illustrated 
by considering diagrams with a  massless $q \bar q$ pair in the final state.
\begin{figure}[t]
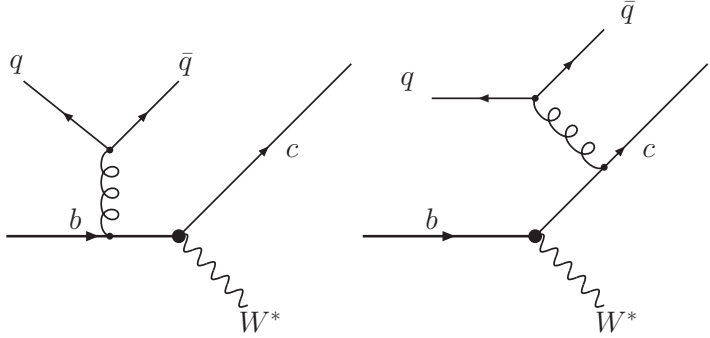

\begin{center}
\includegraphics[angle=0,scale=0.65]{bdec1_fig8.epsi}
\includegraphics[angle=0,scale=0.65]{bdec1_fig9.epsi}
\caption{Diagrams that describe double real emission process with
a massless $q \bar q$ pair in the final state.}   
\label{fig3}
\end{center}
\end{figure}

Consider two  diagrams that 
contribute to the  process $b \to c l \bar \nu_l + q \bar q$, shown in Fig.3.
Upon squaring these diagrams, we find 
that the  gluon splitting $ g^* \to q \bar q$ 
leads to a structure that involves square of the gluon propagator
\be
\frac{ q_{\mu} {\bar q}_{\nu} + q_{\nu} {\bar q}_{\mu} - g_{\mu \nu} q {\bar q}}{(q  \bar q)^2}.
\label{eq:quadr}
\ee
An obvious problem with this result 
is that the singularity associated with the collinear limit is power-like; 
hence, we can not disregard the numerator in the process of 
the sector-decomposition since it provides the  
necessary $q \bar q$ scalar product to 
soften the collinear singularity.  The question is how to parametrize the 
momenta $q,\bar q$  to enable easy extraction of the scalar product 
$q \bar q$ 
from the numerator in Eq.(\ref{eq:quadr}).
To deal with this problem, we need to exploit the fact that, 
when collinear singularity occurs in those diagrams, 
the momenta of $q$ and $\bar q$ become 
parallel to each other. We use this feature  to parametrize the 
scalar products of $q$ and $\bar q$ momenta with, say, the charm quark 
momentum in the following way
\be
c \bar q = \frac{b \bar q}{bq} cq +\sqrt{x_4} x_1 x_2 \Delta_{c \bar q}.
\label{eq55}
\ee
Here $x_4$ describes  the angle between $q$ and $\bar q$, 
$\cos \theta_{12} = 1-2x_4$.
The important point is the factor $\sqrt{x_4}$ in  the 
second term on the right hand side in Eq.(\ref{eq55}); 
it is crucial for regulating collinear 
singularity. On the other hand, the (complicated) 
function $\Delta_{c \bar q}$ does  not 
need to be made explicit in the matrix element to cancel collinear 
singularity. Nevertheless, we give it here for completeness 
\be
\Delta_{c \bar q} = \frac{4 p_c \left ( (m_b - m_c)^2 - m_l^2 
\right ) 
(1-x_3)
\left ( 
\sqrt{x_{4}} \cos \theta_{1c} - \sqrt{1-x_4} \sin \theta_{1c} \cos \phi_c
\right )}{2(m_b -E_c + p_c \cos \theta_{2c}) - E_{g_1} (1 
- \cos \theta_{12})}.
\ee
Very similar manipulations are needed for  scalar products that involve the  
momentum  of the charged lepton and the momenta of $q$ and $\bar q$.

\section{Results}
\label{sect:3}

In this Section, we discuss the  two-loop QCD 
radiative corrections to  $b \to X_c l \bar \nu_l$ transitions 
and present  results for a large number of moments, relevant 
for the experimental analysis and semileptonic fits. 
We begin by writing the decay  rate $b \to X_c l \bar \nu_l$ as
\be
d\Gamma = \frac{G_F^2 \left|V_{cb}\right|^2 m^5_b}{192{\pi}^3}\left(
dF_0+ \frac{\alpha_s}{\pi}dF_1
+ \left(\frac{\alpha_s}{\pi}\right)^2 dF_2
+{\cal O}(\alpha_s^3)
\right),
\label{eq_res_1}
\ee
where ${\rm d}F_i$ stands for the differential decay rate, at leading, next-to-leading and next-to-next-to-leading order, respectively.  The strong
coupling  constant $\alpha_s = \alpha_s(m_b)$
is defined in the ${\overline{\rm MS}}$ scheme in the theory with
three massless flavors; it is renormalized at the value of the $b$-quark mass.
The lepton and hadron 
moments are defined as
\ba
&& L_n\left(E_{\rm cut }\right)
= \frac{\langle{\left(E_l/m_b \right)}^n
\theta\left(E_l-E_{\rm cut } \right)
d\Gamma\rangle}{\langle d\Gamma_0\rangle}, \\
&& H_{ij}\left(E_{\rm cut}\right) =
\frac{\langle \left(({m}_h^2-{m}_c^2)/m_b^2 \right)^i
\left(E_h /m_b \right)^j \theta\left(E_l-E_{\rm cut }\right)
d\Gamma\rangle}{\langle d\Gamma_0\rangle},
\ea
where $E_{l,h}$ are  the lepton and hadron energies in the $b$-quark 
rest frame  and $m_h$ is hadronic invariant mass. Also,  
\be
d\Gamma_0 = \frac{G_F^2 \left|V_{cb}\right|^2 m^5_b}{192{\pi}^3}dF_0,
\ee
and $\langle .. \rangle $ implies integration
of the corresponding quantity over available phase-space\footnote{Available phase space at the parton level is determined by the value of the $b$-quark 
mass. Cuts on the phase-space are shown {\it explicitly} in Eq.(\ref{eq_res_1}).}.
The calculation 
is performed in the pole mass scheme.  For numerical integration, we use 
Vegas \cite{vegas}, implemented in the Cuba library \cite{cuba}. We treat the 
 axial current as suggested in  Ref.\cite{larin}.

The lepton and hadron moments can be computed in an expansion
in the strong coupling constant 
\ba
&& L_n = {L}^{\left(0\right)}_{n} + \frac{\alpha_s}{\pi}L^{\left(1\right)}_n +\left(\frac{\alpha_s}{\pi}\right)^2 \left(\beta_0 L^{\left(2,BLM\right)}_n + L^{\left(2\right)}_n \right)+ ...,
\label{eq341}
\\
&& H_{ij} = {H}^{\left(0\right)}_{ij} + \frac{\alpha_s}{\pi}H^{\left(1\right)}_{ij} +\left(\frac{\alpha_s}{\pi}\right)^2 \left(\beta_0 H^{\left(2,BLM\right)}_{ij} + H^{\left(2\right)}_{ij} \right)+ ...,
\label{eq342}
\ea
where $\beta_0 = 11-2/3n_f$, and $n_f=3$ is the number of quark flavors
that are treated as massless in the computation.
Next-to-leading order and  BLM corrections \cite{blm}  to any kinematic
distribution in  $b \to X_c l \bar \nu_l$ transition are known
\cite{1lrate, diffdistr,trott, kolya1, kolya2}.
Non-BLM corrections $L_n^{(2)},H_{ij}^{(2)}$ for massless lepton 
were computed  very  recently and their detailed investigation is not available.
In the remainder of this Section, we study non-BLM corrections 
to $b \to X_c e \bar \nu_e$ decay in detail. Then we describe results 
for the inclusive rate $b \to X_c \tau \bar \nu_\tau$.

\begin{table}[t]
\begin{center}
\begin{small}
\hspace*{-0.6cm}
\begin{tabular}{|l|r|r|r|r|r|r|r|r|r|r|}
\hline
& $r$ \textbackslash  $\xi$   & 0.0 &0.1  &0.2  &0.3 &0.4 &0.5 &0.6 & 0.7  \\ \hline

${L}_0^{(2)}$&0.2                          &4.01(6)& 3.98(6)& 3.93(7)& 3.73(9)& 3.3(1)& 3.1(1)& 2.47(9)& 2.08(8)\\ \hline                 

${L}_0^{(2)}$&0.22                         &3.74(5)& 3.72(6)& 3.62(7) & 3.48(8)& 3.19(9)& 2.79(9)& 2.34(9)& 1.87(7) \\ \hline                

${L}_0^{(2)}$&0.24                         &3.50(5) & 3.50(5) & 3.35(6) & 3.24(7) & 2.88(8) & 2.57(9) & 2.28(7),& 1.66(6) \\ \hline          

${L}_0^{(2)}$&0.25                         &3.38(4) & 3.37(5) & 3.30(6) & 3.14(7) & 2.84(8) & 2.61(8) & 2.14(7) &1.54(5)  \\ \hline          

${L}_0^{(2)}$&0.26                         &3.27(4) & 3.26(4) & 3.14(5) & 3.03(7) & 2.78(7) & 2.44(7) & 1.95(6) & 1.49(5) \\ \hline                       

${L}_0^{(2)}$&0.28                         &3.05(4) & 3.02(4) & 2.95(5) & 2.84(6) & 2.56(6) & 2.32(7) & 1.80(6) &1.29(4) \\ \hline                    

\hline
\hline
${L}_1^{(2)}$&0.2                          &1.38(2)& 1.38(2)& 1.36(2)& 1.35(2)& 1.30(3)& 1.15(3)& 1.04(3)& 0.90(3)

\\ \hline                 

${L}_1^{(2)}$&0.22                         &1.26(1)& 1.26(1)& 1.25(2)& 1.23(2)& 1.16(2)& 1.12(2)& 0.93(3)& 0.82(3)

 \\ \hline                

${L}_1^{(2)}$&0.24                         &1.16(1)& 1.16(1)& 1.15(1)& 1.14(2)& 1.05(2) &1.00(3) &0.87(2)& 0.71(2) 

\\ \hline          

${L}_1^{(2)}$&0.25                         &1.11(1)& 1.11(1)& 1.09(1)& 1.07(2)& 1.01(2)& 0.93(2)& 0.86(2)& 0.65(2)

  \\ \hline          

${L}_1^{(2)}$&0.26                         &1.06(1) & 1.06(1) & 1.06(1) &1.03(2) &0.99(2) &0.90(2) &0.75(2) &0.61(2)

 \\ \hline                       

${L}_1^{(2)}$&0.28                         &0.97(1)& 0.97(1) &0.96(1) &0.94(1) &0.89(2) &0.83(2)& 0.70(2)

   & 0.53(1)

 \\ \hline                    

\hline
\hline
${L}_2^{(2)}$&0.2                          & 0.514(6)& 0.514(6)& 0.513(6)& 0.508(6)& 0.501(7)& 0.466(9)& 0.45(1)
   & 0.38(1)
\\ \hline                 
${L}_2^{(2)}$&
0.22                         &0.464(5) &0.464(5) &0.462(5) &0.462(6) &0.454(7) &0.422(9) &0.410(9)

   & 0.333(9)

 \\ \hline                

${L}_2^{(2)}$&0.24                         &0.417(4) &0.417(4) &0.416(5) &0.413(5) &0.402(6) &0.377(7) &0.339(8)
   & 0.300(7)           

\\ \hline          

${L}_2^{(2)}$&0.25                         &0.395(4)& 0.395(4) &0.395(4) &0.392(5) &0.376(6) &0.367(7) &0.336(7)  
&0.275(7)
  \\ \hline          

${L}_2^{(2)}$&0.26                         &0.375(4) &0.375(4) &0.374(4) &0.372(4) &0.361(5) &0.345(6) &0.306(7)
   & 0.248(6)                      

 \\ \hline                       
${L}_2^{(2)}$&
0.28                         & 0.336(3) &0.336(3) &0.336(4) &0.331(5) &0.324(6) &0.290(6) &0.274(6)

   & 0.215(6) 

 \\ \hline

\hline
\hline

${L}_3^{(2)}$&0.2                          &0.202(2) &0.202(2) &0.202(2) &0.201(2) &0.197(3) &0.194(3) &0.182(4)
   & 0.159(4)

\\ \hline                 

${L}_3^{(2)}$&0.22                         &0.179(2) &0.179(2) &0.179(2) &0.178(2) &0.177(2) &0.172(3) &0.163(3)

   & 0.139(3)
                      
 \\ \hline                

${L}_3^{(2)}$&0.24                         &0.158(1) & 0.158(2) & 0.158(1) & 0.158(1) & 0.157(2) & 0.151(2) & 0.141(3)

    & 0.120(3)

\\ \hline          

${L}_3^{(2)}$&0.25                         &0.149(1) & 0.149(1) & 0.149(1) & 0.148(1) & 0.147(2) & 0.142(2) & 0.127(3)

    & 0.114(2)

  \\ \hline          

${L}_3^{(2)}$&0.26                         &0.140(1) & 0.140(1) & 0.139(1) & 0.139(1) & 0.138(2) & 0.132(2) & 0.121(2)

    & 0.106(3)

 \\ \hline                       
${L}_3^{(2)}$&
0.28                         &0.123(1) & 0.123(1) & 0.123(1) & 0.122(1) & 0.120(1) & 0.117(2) & 0.107(2)

    & 0.085(2)

  \\ \hline                    
\end{tabular}
\label{tablepton}
\caption{Non-BLM corrections to lepton 
moments $L_i^{\left(2\right)}$ in dependence of $r$ and $\xi$. Vegas integration 
errors are shown in brackets.}
\end{small}
\end{center}
\end{table}

\subsection{Non-BLM corrections and moments of  $b \to X_c e \bar \nu_e$
decays}

In this subsection, we study corrections to semileptonic 
decay  $b \to X_c l \bar \nu_l$, where $l$ is the massless lepton, 
in dependence of the bottom and charm quark masses, 
and  the lepton energy cut.  Because
lepton and hadron moments defined in Eq.(\ref{eq341}) 
and Eq.(\ref{eq342}) are dimensionless,
they depend on the two ratios of the three dimensionfull 
parameters. We choose
$r = m_c/m_b$ and  $\xi=2 E_{\rm cut}/m_b$ as independent variables.
In Table~1  we show non-BLM corrections
to lepton moments $L_{0,1,2,3}$ for a number of $r$ and $\xi$ values.
In Tables~2,3 results for hadron moments 
are given.
These results can, in principle, be used in global fits of semileptonic
decays where $b$ and $c$ masses are {\it parameters} that need to be fitted.

It was observed in \cite{melnikov} that second order QCD corrections
to $b$-decays do not depend  strongly on  kinematics and it is interesting
to further explore this observation.  To this end, we may conjecture that
non-BLM corrections to moments are given by  constant, 
$\xi$-independent  renormalization factors of
the leading order moments. If true, this renormalization factor 
can be determined from  Refs.~\cite{Pak:2008qt,Pak:2008cp}, where 
lepton and hadron energy moments are  analytically computed for 
zero lepton energy cut.  
We will construct  the interpolating function for lepton energy 
moments, following this observation.
At leading order, the lepton energy moments are given by 
\be
{L}_i^{(0)}(r,\xi) = \frac{Y_i(r,\xi)}{Y_0(r,0)},
\ee
where
\be
Y_i(r,\xi) =
\int_{\xi}^{x_m}{\rm d} x
\left[ \frac{2x^2\left(x_m-x\right)^2}{\left(1-x\right)^3}\right]\\
\left(6-6x+xx_m+2x^2-3x_m\right)\left(\frac{x}{2}\right)^i.
\label{eq898}
\ee
In Eq.(\ref{eq898}) $x_m= 1-r^2$ and the integration 
is over $x = 2E_l/m_b$.
Since  the integration here is elementary but the resulting formulas
are lengthy, we do not present the  results of the integration 
 here. We introduce the interpolating function  by defining
\be
L_{i}^{(2),\rm in}(r,\xi) =
\frac{L_{i}^{(2)}(r,0)}{L_{i}^{(0)}(r,0)} L_{i}^{(0)}(r,\xi),
\label{eq12}
\ee
so that the normalization of the non-BLM correction to the moment is fixed by
its value at zero lepton energy cut
and the shape is taken to coincide with the  leading order shape.  The
interpolated  moments $L_{i}^{(2),\rm in}(r,\xi)$ are given in Table~4.
Comparing computed and interpolated  moments, we observe that 
$L_{i}^{(2),\rm in}(r,\xi)$ provides excellent approximation 
to $L_{i}^{(2)}(r,\xi)$ for 
small values of $\xi$. 
However, the agreement becomes progressively worse for larger values of $\xi$.
For example, a typical deviation between the interpolated  and the explicitly
computed non-BLM moments
for $\xi =0.7$ and $m_c/m_b = 0.28$ can be as much as twenty
percent.  Finally, we point out that a very similar behavior is observed 
for non-BLM  hadron  energy moments.

\begin{table}[t]
\begin{center}
\label{tabhade}
\begin{small}
\hspace*{-0.6cm}
\begin{tabular}{|l|r|r|r|r|r|r|r|r|r|r|}
\hline
& $r$ \textbackslash  $\xi$   & 0.0 &0.1  &0.2  &0.3 &0.4 &0.5 &0.6 & 0.7  \\ \hline

${H}_{01}^{(2)}$&0.2                        & 1.39(3)& 1.38(3)& 1.32(4)& 1.28(5)& 1.19(5)& 1.19(5)& 0.94(4)& 0.78(3)

\\ \hline                 

${H}_{01}^{(2)}$&0.22                       & 1.34(3)& 1.33(3)& 1.32(3)& 1.24(4)& 1.16(4)& 1.08(4)& 0.85(4)& 0.72(3)

 \\ \hline                

${H}_{01}^{(2)}$&0.24                       & 1.29(2)& 1.28(3)& 1.23(3)& 1.15(3)& 1.09(4)& 1.02(4)& 0.88(3)& 0.65(3)

\\ \hline          

${H}_{01}^{(2)}$&0.25       &1.27(2)& 1.26(2)& 1.21(3)& 1.18(3)& 1.05(3)& 0.95(4)& 0.76(3)& 0.64(2)

  \\ \hline          

${H}_{01}^{(2)}$&0.26               & 1.24(2)& 1.23(2)& 1.18(3)& 1.11(3)& 1.00(3)& 0.95(3)& 0.76(3)& 0.60(2)

 \\ \hline                       

${H}_{01}^{(2)}$&0.28                 & 1.20(2)& 1.19(2)& 1.15(3)& 1.12(3)& 0.99(3)& 0.89(3)& 0.69(2)& 0.54(2)\\

\hline
\hline

${H}_{02}^{(2)}$&0.2                          &0.46(1)& 0.46(1)& 0.47(2)& 0.42(2)& 0.38(2)& 0.39(2)&
    0.35(2)& 0.28(1)

\\ \hline                 

${H}_{02}^{(2)}$&0.22                         &0.46(1)& 0.46(1)& 0.44(2)& 0.45(2)& 0.39(2)& 0.39(2)

&    0.31(2)& 0.27(1)

 \\ \hline                

${H}_{02}^{(2)}$&0.24                         &0.46(1)& 0.46(1)& 0.45(1)& 0.44(2)& 0.38(2)& 0.35(2)
&    0.33(1)& 0.26(1)

\\ \hline          

${H}_{02}^{(2)}$&0.25                         &0.46(1)& 0.46(1)& 0.45(1)& 0.45(2)& 0.41(2)& 0.35(2)&    0.34(1)& 0.25(1)

  \\ \hline          

${H}_{02}^{(2)}$&0.26                         &0.46(1) &0.45(1)& 0.44(1)& 0.42(2)& 0.38(2)& 0.37(2)&    0.32(1)& 0.24(1)

 \\ \hline                       

${H}_{02}^{(2)}$&0.28                         &0.46(1)& 0.46(1)& 0.45(1)& 0.44(2)& 0.39(2)& 0.33(1)&    0.30(1)& 0.24(1)

\\

\hline
\hline

${H}_{03}^{(2)}$&0.2                          &0.140(7)& 0.137(8)& 0.140(9)& 0.12(1)& 0.14(1)& 0.14(1)&   0.122(8)& 0.104(6)

\\ \hline                 

${H}_{03}^{(2)}$&0.22                         &0.146(6)& 0.143(7)& 0.142(8)& 0.152(9)& 0.13(1)& 0.125(9)&   0.123(7)& 0.107(5)

 \\ \hline                

${H}_{03}^{(2)}$&0.24                         &0.152(6)& 0.150(6)& 0.143(7)& 0.145(9)& 0.124(9)& 0.132(8)&  0.133(8)& 0.095(5)

\\ \hline          

${H}_{03}^{(2)}$&0.25                         &0.155(5)& 0.153(6)& 0.158(7)& 0.145(8)& 0.150(8)& 0.133(8)&   0.124(7)& 0.103(4)

  \\ \hline          

${H}_{03}^{(2)}$&0.26                         &0.158(5)& 0.157(6)& 0.155(7)& 0.152(8)& 0.140(9)& 0.145(7)&    0.130(6)& 0.101(4)

 \\ \hline                       

${H}_{03}^{(2)}$&0.28                         &0.164(5)& 0.162(6)& 0.158(7)& 0.164(8)& 0.153(7)& 0.128(7)&   0.125(6)& 0.096(4)
\\

\hline
   
\end{tabular}
\caption{ Non-BLM corrections to hadron energy moments $H_{0i}^{(2)}$ 
in dependence of $r$ and $\xi$. Vegas integration errors are shown 
in brackets.}
\end{small}
\end{center}
\end{table}

Note that by increasing the cut on the lepton energy, the phase-space
is restricted to the region where soft gluon radiation becomes relatively more
important and, hence, the dynamics of the final state changes with the
increase of the cut on the lepton energy. It is therefore clear that 
for moments defined in Eqs.(\ref{eq341},\ref{eq342}) perturbative corrections 
at different lepton energy cuts are {\it not 
correlated}, i.e. different physics becomes important for different values
of the lepton energy cut. On the other hand, we point out that the moments measured in 
experimental analysis correspond to ratios of $L$-moments defined in 
Eqs.(\ref{eq341},\ref{eq342}); as we explain below, this difference is 
essential for 
understanding importance of QCD radiative corrections in the global fits. 

We turn to the discussion of the potential impact 
that computed corrections may have  on the extraction of fundamental quantities 
in heavy quark physics, such as 
$|V_{cb}|, m_b, m_c, \mu_\pi^2, \,\mu_G^2$ etc. from 
global fits to semileptonic moments.  We stress that we do not attempt to
perform a fit to  data on semileptonic 
moments, leaving this task to experts.  However,  we find it instructive 
to illustrate shifts that may be expected in the values of, 
e.g. the $|V_{cb}|$ and the $b$-quark mass,
if non-BLM corrections are included.

\begin{table}[tb]
\begin{center}
\begin{small}
\label{tabhinvmass}
\hspace*{-0.6cm}
\begin{tabular}{|l|r|r|r|r|r|r|r|r|r|r|}

\hline

 & $r$ \textbackslash   $\xi$   & 0.0 &0.1  &0.2  &0.3 &0.4 &0.5 &0.6 & 0.7  \\ \hline

${H}_{10}^{(2)}$&0.2                &-0.443& -0.441& -0.423& -0.387& -0.331& -0.259

&    -0.181& -0.107(1)

\\ \hline                 

${H}_{10}^{(2)}$&0.22               &-0.402& -0.399& -0.383& -0.347& -0.294& -0.228&

    -0.156& -0.090(1)

 \\ \hline                

${H}_{10}^{(2)}$&0.24               &-0.364& -0.361& -0.346& -0.313& -0.262& -0.200&

    -0.134& -0.0739

\\ \hline          

${H}_{10}^{(2)}$&0.25               &-0.346&-0.343 &-0.328& -0.296& -0.246& -0.187&

    -0.124& -0.0662

  \\ \hline          

${H}_{10}^{(2)}$&0.26               &-0.329 &-0.326& -0.311& -0.280& -0.231& -0.174

   & -0.114& -0.0602

 \\ \hline                       

${H}_{10}^{(2)}$&0.28               &-0.297& -0.295& -0.280& -0.250& -0.205& -0.151&

    -0.0966& -0.0484(4)

\\

\hline   
\hline

${H}_{11}^{(2)}$&0.2               & -0.215& -0.213& -0.204& -0.184& -0.155& -0.120(1)& -0.082(1)& -0.0479(4)

\\ \hline                 

${H}_{11}^{(2)}$&0.22              &-0.198& -0.197& -0.188& -0.169& -0.141& -0.107(1)&  -0.0721& -0.0408(3)

 \\ \hline                

${H}_{11}^{(2)}$&0.24              &-0.183& -0.181& -0.173& -0.154& -0.128& -0.0956&  -0.0632& -0.0342(3)

\\ \hline          

${H}_{11}^{(2)}$&0.25              &-0.175& -0.174& -0.165& -0.147& -0.121(1)& -0.0900& -0.0590& -0.0313

  \\ \hline          

${H}_{11}^{(2)}$&0.26              &-0.168& -0.167& -0.158& -0.141& -0.114(1)& -0.0848& -0.0548& -0.0284

 \\ \hline                       

${H}_{11}^{(2)}$&0.28              &-0.155& -0.153& -0.145& -0.128& -0.103(1)& -0.0748& -0.0469& -0.0232(2)

\\

\hline   
\hline

${H}_{12}^{(2)}$&0.2                & -0.107& -0.106& -0.100& -0.090& -0.0747& -0.0565&-0.0383 &-0.0217

\\ \hline                 

${H}_{12}^{(2)}$&0.22               &-0.0998& -0.0989& -0.0937& -0.0837& -0.0688& -0.0513& -0.0341& -0.0189

 \\ \hline                

${H}_{12}^{(2)}$&0.24               &-0.0935& -0.0926& -0.0877& -0.0773& -0.0629&    -0.0468& -0.0304& -0.0161

\\ \hline          

${H}_{12}^{(2)}$&0.25               &-0.0905& -0.0896& -0.0846& -0.0748& -0.0604& -0.0443& -0.0285& -0.0148

  \\ \hline          

${H}_{12}^{(2)}$&0.26               &-0.0876& -0.0867& -0.0817& -0.0719& -0.0579&  -0.0423& -0.0267& -0.0136

 \\ \hline                       

${H}_{12}^{(2)}$&0.28               &-0.0820& -0.0810& -0.0761& -0.0664& -0.0531& -0.0377& -0.0234& -0.0113

\\

\hline
\end{tabular}
\caption{Non-BLM corrections to hadron invariant mass 
moments ${H}_{1j}^{(2)}$ in dependence of  $r$ and $\xi$.
Vegas integration errors are shown 
in brackets.  Integration errors are not shown if  they are significantly smaller than one percent.}
\end{small}
\end{center}
\end{table}

\begin{figure}[t]
\begin{center}
\includegraphics[angle=-90,scale=0.4]{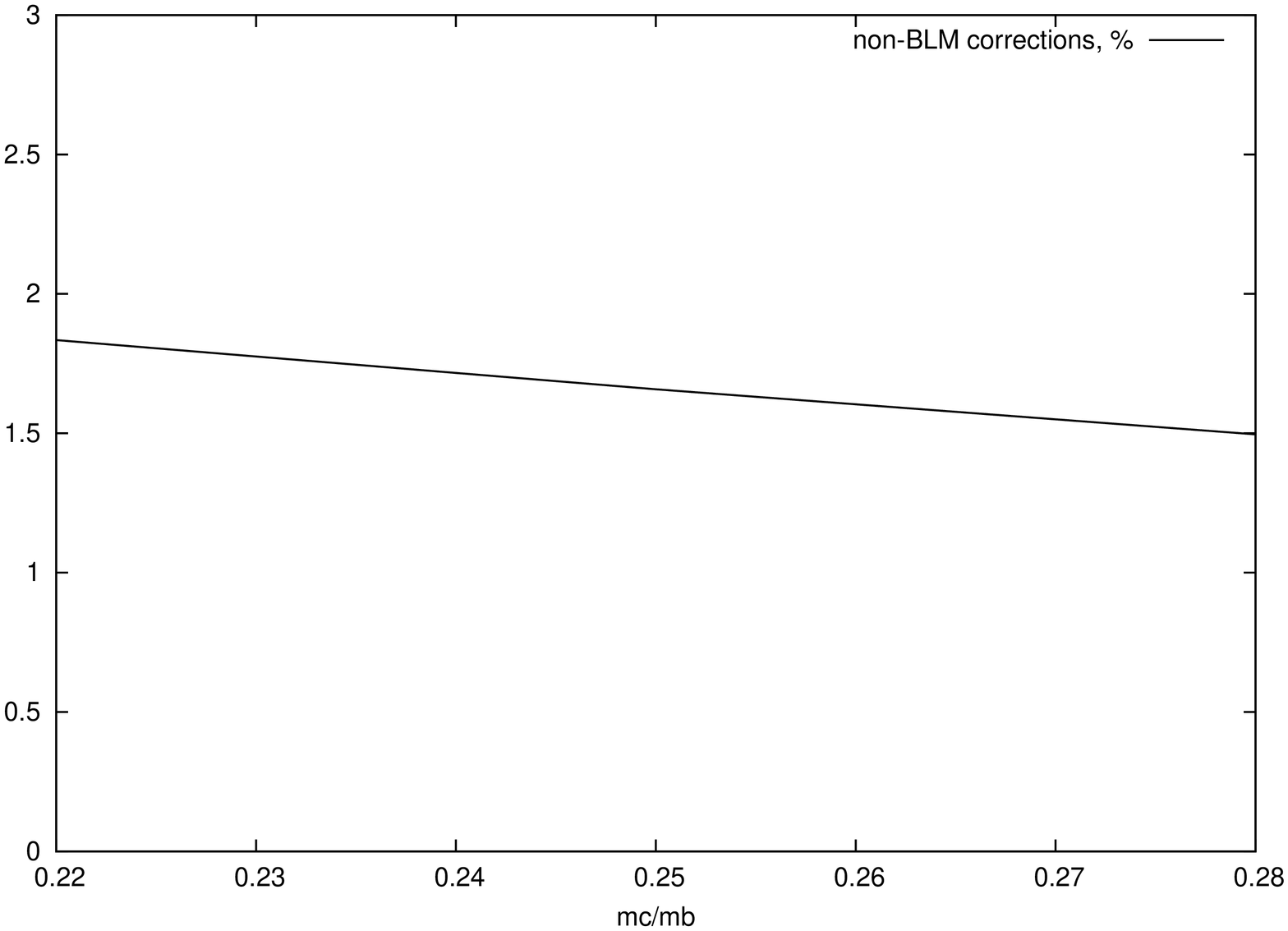}
\vspace*{0.5cm}
\caption{
Non-BLM correction  to the total rate, in percent, 
 for zero lepton energy cut, as 
a function of the charm quark mass to bottom quark mass ratio.
We use $\alpha_s = 0.22$.}
\label{fig25}
\end{center}
\end{figure}

We begin with the discussion of the CKM matrix element $|V_{cb}|$. 
Since the $|V_{cb}|$ is obtained  from the normalization of the 
partial decay rate, it is mostly sensitive to 
QCD corrections to the moment $L_{0}$. The non-BLM corrections 
to that moment for $\xi = 0$ and various values of $r$ are shown 
in Fig.\ref{fig25}. We see that, for realistic ratios of quark masses, 
the non-BLM corrections to $L_0$ are between $2$ and $1.5$ percent. 
Since experimental measurement fixes $|V_{cb}|^2 L_0$, one can 
 expect that $|V_{cb}|$ changes by about $-1$ percent, when non-BLM corrections 
are included. This is compatible with the uncertainty 
in  $|V_{cb}|$ as currently 
estimated.

BABAR collaboration measured a number of lepton energy moments 
for $b \to c l \bar \nu_l $
transitions  with high precision \cite{babar2009}. For the 
illustration, we use
their measurement of  the moment ${\cal M}_1$ for two values of the lepton energy cut
\be
{\cal M}_1(E_{\rm cut}) =
\frac {\int \limits_{E_{\rm cut}}^{} E_l \;  {\rm d} \Gamma}{\int \limits_{E_{\rm cut}}^{} {\rm d} \Gamma}
\;\; =  \left \{
\begin{array}{cc}
1437.6(4.0)(5.7)~{\rm MeV}, & E_{\rm cut} = 0.6~{\rm GeV}; \\
1773.7(1.9)(1.1)~{\rm MeV}, & E_{\rm cut} = 1.5~{\rm GeV}.
\end{array}
\right.
\label{eqr:3}
\ee
These (and other) results are used to extract the following 
values of the
bottom and charm quark masses in the kinetic scheme \cite{kin}
$m_b^{\rm fit} = 4.55(5)~{\rm GeV}$ and
$m_c^{\rm fit} = 1.08(7)~{\rm GeV}$,
where we combined all uncertainties in quadratures.  In terms of 
lepton moments computed in this paper, and neglecting non-perturbative
contributions, we find  
\be
{\cal M}_1(E_{\rm cut}) = m_b {\cal M}_{1}^{\rm pt},\;\;\;\;\;
{\cal M}_{1}^{\rm pt} = \frac{L_1(E_{l \rm cut})}{L_0(E_{\rm cut})}.
\label{eqr:4}
\ee

\begin{table}[h] 
\begin{center}
\label{tableptonfit}
\begin{tabular}{|l|r|r|r|r|r|r|r|r|r|r|}
\hline
& $r$ \textbackslash  $\xi$ & 0    &0.1  &0.2  &0.3 &0.4 &0.5 &0.6 & 0.7  \\ \hline

${L}_0^{(2)\rm in}$&0.2      &4.01                    & 4.00& 3.94 & 3.77 & 3.48 & 3.05& 2.48 & 1.77\\ \hline                 

${L}_0^{(2)\rm in}$&0.22     &3.74                    & 3.74 & 3.68 & 3.51 & 3.23 & 2.80& 2.25 & 1.57 \\ \hline                

${L}_0^{(2)\rm in}$&0.24     &3.50                    & 3.49  & 3.43  & 3.27 & 2.99 & 2.58  & 2.04& 1.38 \\ \hline          

${L}_0^{(2)\rm in}$&0.25     &3.38                    & 3.37 & 3.30 & 3.15 & 2.88 & 2.47 & 1.93 &1.29  \\ \hline          

${L}_0^{(2)\rm in}$&0.26     &3.27                    & 3.26 & 3.19 & 3.05 & 2.77 & 2.36 & 1.83 & 1.20 \\ \hline                       

${L}_0^{(2)\rm in}$&0.28     &3.05                    & 3.04 & 2.97 & 2.83 & 2.56 & 2.16 & 1.64 &1.03 \\ \hline                    

\hline
\hline
${L}_1^{(2)\rm in}$&0.2      &1.38                     &1.36& 1.36& 1.33& 1.28& 1.18& 1.01& 0.77

\\ \hline                 
${L}_1^{(2)\rm in}$&0.22     &1.26                     &1.26& 1.25& 1.23& 1.18& 1.08& 0.91& 0.68
 \\ \hline                
${L}_1^{(2)\rm in}$&0.24     &1.16                     &1.16& 1.15& 1.13& 1.08& 0.98& 0.82& 0.59
\\ \hline          
${L}_1^{(2)\rm in}$&0.25     &1.11                     &1.11& 1.10& 1.08& 1.03& 0.93& 0.77& 0.55 
  \\ \hline          
${L}_1^{(2)\rm in}$&0.26     &1.06                     &1.06& 1.05&    1.03&    0.98&  0.88&0.73& 0.51
 \\ \hline                       
${L}_1^{(2)\rm in}$&0.28     &0.97                     &0.97 &    0.96& 0.94&     0.89& 0.80& 0.65&0.43
 \\ 

\hline
\hline
${L}_2^{(2)\rm in}$&0.2    &0.514                                &0.514 &0.514& 0.510& 0.500& 0.475& 0.425& 0.339
        \\ \hline                 

${L}_2^{(2)\rm in}$&0.22    &0.464                           &0.464& 0.464& 0.461& 0.450& 0.426& 0.377& 0.295

 \\ \hline                

${L}_2^{(2)\rm in}$&0.24      &0.417                         &0.417& 0.417& 0.414& 0.404& 0.380& 0.332& 0.253 

\\ \hline          

${L}_2^{(2)\rm in}$&0.25        &0.395                       &0.395& 0.395& 0.392& 0.382& 0.358& 0.312& 0.234
  \\ \hline          

${L}_2^{(2)\rm in}$&0.26        &0.375                       &0.375& 0.375& 0.372& 0.362& 0.338& 0.292& 0.216

 \\ \hline                       

${L}_2^{(2)\rm in}$&0.28         &0.336                      &0.336& 0.336& 0.333& 0.323& 0.300& 0.255& 0.182
 \\ 

\hline
\hline

${L}_3^{(2)\rm in}$&0.2           &0.202                     &0.202& 0.202& 0.202& 0.200& 0.193& 0.179& 0.149
        
\\ \hline                 

${L}_3^{(2)\rm in}$&0.22          &0.179                     &0.179& 0.179& 0.179& 0.177& 0.171& 0.156& 0.128

 \\ \hline                

${L}_3^{(2)\rm in}$&0.24           &0.158                    &0.158& 0.158& 0.158& 0.156& 0.150& 0.136& 0.109

\\ \hline          

${L}_3^{(2)\rm in}$&0.25           &0.149                    &0.149& 0.149& 0.149& 0.147& 0.141& 0.127& 0.100
  \\ \hline          

${L}_3^{(2)\rm in}$&0.26           &0.140                    &0.140& 0.140& 0.140& 0.138& 0.132& 0.118& 0.092

 \\ \hline                       

${L}_3^{(2)\rm in}$&0.28           &0.123                    &0.123& 0.123& 0.123& 0.121& 0.115& 0.102& 0.077

 \\ \hline                    
\end{tabular}
\caption{Moment $L_i^{\left(2\right) \rm in}$ in dependence 
of $r$ and $\xi$. Entries at $\xi=0$ are initial 
conditions for the fit; see text for details.}
\end{center}
\end{table}

The non-BLM corrections to lepton moments were not accounted for in \cite{babar2009}.
To estimate their impact, we compute $L_0$ and $L_1$ for
$m_b = 4.55(5)~{\rm GeV}$
and $m_c = 1.08(7)~{\rm GeV}$ and $E_{\rm cut} = 0.6~{\rm and}~1.5~{\rm GeV}$.
The results are given in Table~5.
Expanding Eq.(\ref{eqr:4}) around $m_b = m_b^{\rm fit} + \delta m_b$, to
account for the shift in the $b$-quark mass, induced by including
non-BLM corrections in the calculation of  ${\cal M}_{1}^{\rm pt}$, 
we find
\be
\delta m_b = -\frac{\left ( m_b^{\rm fit} \right ) ^2 
\; \delta {\cal M}_{1}^{\rm pt}}{{\cal M}_1
+ m_b^{\rm fit} \frac{{\rm d}{\cal M}_{1}^{\rm pt}}{{\rm d} m_b}}.
\label{eqshift}
\ee
In Eq.(\ref{eqshift}), $\delta {\cal M}_{1}^{\rm pt}$ 
accounts for the change in the $b$-quark mass  due to
${\cal O}(\alpha_s^2)$
non-BLM corrections in ${\cal M}_{1}^{\rm pt}$. 
Explicitly, 
\be
\delta {\cal M}_{1}^{\rm pt} =
\left (\frac{\alpha_s}{\pi} \right )^2 
\left [   
\frac{L_{1}^{(2)}}{L_{0}^{(0)}}
-\frac{L_{1}^{(0)}L_{0}^{(2)}}{\left ( L_{0}^{(0)} \right )^2}
-\frac{L_{0}^{(1)}L_{1}^{(1)}}{\left ( L_{0}^{(0)} \right )^2}
+\frac{L_{1}^{(0)}\left(L_{0}^{(1)}\right)^2}{\left ( L_{0}^{(0)} \right )^3}
\right ].
\ee
To calculate ${\rm d}{\cal M}_{1}^{\rm pt}/{\rm d}m_b$, we employ the leading
order expression for ${\cal M}_{1}^{\rm pt}$, 
neglecting both perturbative
and non-perturbative corrections; the impact of this derivative on the 
shift in the $b$-quark mass is small.  We find the following change in the 
value of the $b$-quark {\it pole mass}
\be
\delta m_b =
\left \{
\begin{array}{cc}
-6.6(1.6)
~{\rm MeV}, & E_{\rm cut} = 0.6~{\rm GeV}; \\
-6.4(4.0)~{\rm MeV}, & E_{\rm cut} = 1.5~{\rm GeV},
\end{array} \right.
\label{eq:shift}
\ee
where we used $\alpha_s(m_b) = 0.22$.  In brackets,  the uncertainties in the mass 
shift related to numerical integration errors are indicated. Note a very strong 
amplification of the numerical integration errors when we pass from $L$ to ${\cal M}$ 
moments -- integration errors are just a few percent in the former and up to 
$60$ percent  in the latter. This implies a very strongly cancellation between 
radiative corrections in the ratio of $L_{1}$ and $L_0$.

\begin{table}[t]
\begin{center}
\label{tabmom}
\vspace{0.1cm}
\begin{tabular}{|c|c|c|c|c|}
\hline\hline
$n$ & $E_{cut}$, GeV & $L_n^{(0)}$ & $L_n^{(1)}$ & $L_n^{(2)}$ \\ \hline\hline
$0$ & $0.6$ &  $0.9552269$&   $-1.723893$        &  $3.29(6)$   \\ \hline
$1$ & $0.6$ &  $0.3065502$ & $-0.559955$       &  $1.16(1)$   \\ \hline
$0$ & $1.5$ &  $0.4790680$ & $-0.881472$       &  $1.97(6)$   \\ \hline
$1$ & $1.5$ &  $0.1871146$&  $-0.350026$       &  $0.83(2)$   \\ \hline
\end{tabular}
\caption{Lepton energy moments for $m_b = 4.55~{\rm GeV}$
and $m_c = 1.08~{\rm GeV}$.}
\end{center}
\end{table}

Note that Eq.(\ref{eq:shift}) gives corrections in the pole mass scheme and that 
additional non-BLM corrections appear if the pole mass is transformed 
to the kinetic mass \cite{kin}; those corrections 
were computed in \cite{czmeu}.   For the kinetic mass at $\mu= 1~{\rm GeV}$, the
additional shift  is  about $15~{\rm MeV}$, so that the total shift 
\be
\delta m_b^{\rm kin}(1~{\rm GeV}) \approx  10~{\rm MeV}
\label{eq457}
\ee
can be expected\footnote{We point out that explicit  formulas
that relate perturbative QCD corrections to the inclusive semileptonic 
$b \to X_c l \bar \nu_l$ decay width  in the pole and kinetic schemes are given 
in Ref.~\cite{imp}.}. There are two ways to look 
at the significance of this result. We can 
compare it to the uncertainty in the $b$-quark mass
of about $40-50~{\rm MeV}$, typically
obtained in  fits to moments of  
semileptonic $b$-decays \cite{fit1,fit1a,fit2,babar2009}. This comparison 
indicates that the shift shown in Eq.(\ref{eq457}) is rather small.  
On the other hand, the error on the $b$-quark mass in the fits is related 
to the fact that global fits are not very sensitive to $m_{b}$ and $m_c$ 
individually; rather, the linear combination $m_b - 0.6 m_c$ is restricted 
to about $6~{\rm MeV}$.  Because we estimated the shift in the 
$b$-quark mass for the fixed value of the charm quark mass, 
a more relevant uncertainty in the $b$-quark 
mass to compare should be just these $6~{\rm MeV}$, which is  similar
to our estimate of $\delta m_b$ due to  non-BLM corrections\footnote{We thank 
N.~Uraltsev for emphasizing this point to us.}.
We emphasize  that the change in the $b$-quark mass shown 
in Eq.(\ref{eq457}) is only an 
 estimate and  more careful calculation, that 
includes larger number of moments and  
simultaneous extraction of  all heavy quark 
parameters, is required. 

Finally, we stress that, regardless of what uncertainty $\delta m_b$ should 
be compared to, Eq.(\ref{eq:shift}) is remarkable since it shows 
that the correction  to a  low-energy observable due to 
two-loop non-BLM QCD 
effects is {\it much smaller} than the naive estimate suggests 
\be
\frac{\delta m_b}{m_b} \sim 10^{-3} \ll C_F C_A 
\left ( \frac{\alpha_s}{\pi} \right )^2 \sim 
2 \cdot 10^{-2}.
\ee
This feature is a consequence of a 
very strong cancellation between corrections to $L^{(1)}$ and
$L^{(0)}$, when the ratio of the two is taken to compute
${\cal M}_{1}^{\rm pt}$. 
To illustrate this point, note that
if we set non-BLM corrections in
$L^{(0)}$ to zero, the shift $\delta m_b$ increases
from about $-7~{\rm MeV}$ , as shown in
Eq.(\ref{eq:shift}), to about $-100~{\rm MeV}$.  It appears therefore that 
high degree of cancellations of radiative corrections between
different $L$-moments is crucial for claiming very small errors
in $m_b$, $m_c$ etc.  In this respect, 
it is important 
to understand the origin of these cancellations since there are yet 
higher-order perturbative effects about which 
nothing is known at present and that, naively, 
are of the same order of magnitude as the non-BLM corrections computed 
in this paper.  For example, although  leading order BLM 
corrections  ${\cal O}(\alpha_s^n \beta_0^{n-1})$ are known and resummed
\cite{kolya2},
subleading BLM  effects ${\cal O}(\alpha_s^n \beta_0^{n-2})$ are
not known beyond $n=2$. But, because $\beta_0 \sim 10 $ is large,
one should expect that three-loop subleading BLM corrections 
to $L_0$ and $L_1$ 
are of the same order of magnitude as the two-loop non-BLM effects considered 
in this paper
$\alpha_s^3 \beta_0 \sim \alpha_s^2$. 
The only way to avoid
large shifts in the $b$-quark mass  is to have nearly complete cancellation
between these three- and higher-loop corrections
to $L_0$ and $L_1$. The {\it degree} of such cancellation is an assumption
in existing fits to semileptonic moments in $B$-decays, as long as the 
origin of this cancellation is not clearly understood. To this end, 
it is interesting   to give a few arguments in favor of non-accidental 
nature of these cancellations.

For example, it is easy to see that 
in the limit of a very high cut on the lepton energy, 
all perturbative corrections to normalized moments vanish. Indeed, 
we consider the $n$-th 
normalized moment of the lepton energy, computed in perturbation theory
\be
{\cal M}_n^{\rm pt}(E_{\rm cut}) = \frac{L_n(E_{\rm cut})}{L_0(E_{\rm cut})}.
\ee
We now make a simple observation that 
\be
\lim_{E_{\rm cut} \to E_{l}^{\rm max}} {\cal M}_n^{\rm pt}(E_{\rm cut}) = \left ( E_{l}^{\rm max} \right )^n
= \frac{(m_b^2-m_c^2)^n}{(2 m_b)^n}, 
\label{eq765}
\ee
independent of the strong coupling constant $\alpha_s$. 
Eq.(\ref{eq765}) 
implies {\it perfect} cancellation of all 
radiative corrections to normalized  moments in that limit.
Corrections  to this result scale as 
${\cal O}(n \alpha_s^k (E_{l}^{\rm max} - E_{\rm cut})/E_{l}^{\rm max} 
 \ln^j((E_{l}^{\rm max} - E_{\rm cut})/m_b) $; 
they are clearly much smaller than the naive ${\cal O}(\alpha_s^k)$ 
estimate of a  $k$-loop QCD corrections.

Moreover, one can relax the requirement of a high lepton 
energy cut, by making the 
following observations. The lepton energy distribution has a 
peak, at $ E_l \approx 0.8 E_{l}^{\rm max}$. In the limit when 
this peak is infinitely narrow, {\it normalized} moments of, say, lepton 
energy are obviously 
protected from radiative corrections. 
Hence, deviations from the ``no radiative-corrections'' limit must be 
correlated with the broadness of the peak. To this end, note that 
the peak appears to be fairly narrow 
-- for example, the position of the peak is 
{\it only}  fifteen to twenty percent higher than the average 
lepton energy. We believe 
that results of explicit computations supplemented with these 
considerations, give a strong argument in favor of non-accidental nature 
of the observed cancellations in normalized moments for any value 
of the cut on the lepton energy and suggest that 
similar  cancellations persist to  higher 
orders in  perturbative QCD.

\subsection{Decay $B \to X_c \tau \bar \nu_\tau$}

We come to the discussion of the NNLO QCD corrections to semileptonic
$B$ decays
into final states with the charm quark and the $\tau$ lepton.  The corresponding
branching ratios were measured at LEP by ALEPH and
OPAL collaborations \cite{expaleph,expopal}. The results read
\be
{\rm Br}(B \to X_c \tau \bar \nu_\tau)
= 
\left \{
\begin{array}{cc}
(2.43 \pm 0.32) \times 10^{-2}, & {\rm ALEPH}, \\ 
(2.78 \pm 0.54) \times 10^{-2}, & {\rm OPAL},
\end{array} 
\right.
\ee
where we added statistical and systematic errors in quadratures.
We employ the ALEPH measurement in the following
numerical computation. Using the world average  for semileptonic
branching ratio into the massless lepton,
\be
{\rm Br}(B \to X_c l \bar \nu_l ) = (10.25 \pm 0.25) \times 10^{-2},
\ee
we find the ratio of the two branching fractions
\be
{\cal R} = \frac{\Gamma(B \to X_c \tau \bar \nu_\tau)}{\Gamma(B \to X_c l \bar \nu_l)}
\approx 0.237(31).
\label{eq543}
\ee

The ratio ${\cal R}$ can be very accurately predicted in perturbative QCD.
Indeed, setting $m_b = 4.6~{\rm GeV}$, $m_c = 1.15~{\rm GeV}$ and
$m_\tau = 1.8~{\rm GeV}$, we obtain  the following results
for  semileptonic decay rates
\ba
&& \frac{\Gamma(b \to c e \bar \nu_e)}{\Gamma_0} = 
z_0\left (\rho_c, \rho_e\right )
\left[1 + \frac{\alpha_s}{\pi}\left(-1.777\right)+
\left(\frac{\alpha_s}{\pi}\right)^2
\left( -1.92\beta_0 + 3.38\right)\right],\\
&& \frac{\Gamma(b \to c \tau \bar \nu_\tau)}{\Gamma_0} =
z_0\left (\rho_c,\rho_\tau \right )
\left[1 + \frac{\alpha_s}{\pi}\left(-1.462\right)+
\left(\frac{\alpha_s}{\pi}\right)^2
\left( -1.82\beta_0 + 3.16\right)\right],
\ea
where $\rho_c = m_c^2/m_b^2, \rho_e = 0, \rho_\tau = m_\tau^2/m_b^2$ and 
$\Gamma_0 = G_F^2 |V_{cb}|^2 m_b^5/(192\pi^3)$.
The function $z_0(\rho_q,\rho_l)$ reads \cite{koyrakh,neubert,Balk:1993sz}
\ba
 z_0(\rho_q,\rho_l) 
&& = \sqrt{\lambda} 
\left ( 1 - 7 \rho_q - 7 \rho_q^2+\rho_q^3
- 7 \rho_l - 7 \rho_l^2 + \rho_l^3
+ \rho_q \rho_l \left (12 - 7 \rho_q - 7 \rho_l \right ) \right )
\nonumber \\
&& 
+ 12 \rho_q^2(1-\rho_l)^2 \log \frac{1+v_q}{1-v_q}
+ 12 \rho_l^2(1-\rho_q)^2 \log \frac{1+v_l}{1-v_l},
\ea
where $\lambda(\rho_q,\rho_l) = 
1 + \rho_q^2 + \rho_l^2 - 2\rho_q - 2\rho_l - 2\rho_q \rho_l$ and 
$\displaystyle v_q = \frac{\sqrt{\lambda}}{1 + \rho_q - \rho_l}$, 
$\displaystyle v_l = \frac{\sqrt{\lambda}}{1 + \rho_l - \rho_q}$.

Taking the ratio, we find
\begin{eqnarray}
{\cal R}^{\rm pert}&&  = \frac{z_0(\rho_c,\rho_\tau)}{z_0(\rho_c,\rho_e)}
\left ( 1 + 0.315 \left ( \frac{\alpha_s}{\pi} \right )  + \left (0.9_{\rm BLM} + 0.34 \right )
\left ( \frac{\alpha_s}{\pi} \right )^2 \right) 
\nonumber  \\
&& = \frac{z_0(\rho_c,\rho_\tau)}{z_0(\rho_c,\rho_e)}
\left ( 1 + 0.0221_{{\cal O}(\alpha_s)} + 0.0044_{\rm BLM} +
0.0017_{\rm non-BLM} \right),
\label{eq:r8}
\end{eqnarray} 
where at the last step  $\alpha_s = 0.22$ was used. We observe 
that  QCD effects in the ${\cal R}$ ratio
are very small.  We stress that while the ratio of leading order
decay rates is a rapidly changing function of $m_b, m_c$ and
$m_\tau$, radiative corrections to ${\rm Br}(b \to X_c \tau \bar \nu_\tau )$
and ${\rm Br}(b \to X_c l \bar \nu_l )$ are correlated, so that they
cancel out in the ratio largely independent of the quark masses.
We point out that
non-perturbative corrections to the ${\cal R}$ ratio were computed
in \cite{koyrakh,neubert,Balk:1993sz} and were found to be of the order of 
minus four  percent. Interestingly, not only perturbative and non-perturbative 
corrections are small {\it individually}, but they also tend to cancel each other.
Given that perturbative and non-perturbative effects in the ${\cal R}$ ratio are very small,  we  fix the $b$-quark mass to its value from 
semileptonic fits $m_b = 4.55~{\rm GeV}$ and 
require 
\be
\frac{z_0(\rho_c,\rho_\tau)}{z_0(\rho_c,\rho_e)}
  = 0.237(31),
\label{eq:constr}
\ee 
to determine the charm quark mass. 
The dependence of decay rates on quark masses
at leading order is well-known; it can be extracted from
Refs.\cite{koyrakh,neubert,Balk:1993sz}. We obtain
\be
m_c = 1.040(200)~{\rm GeV}.
\ee
This $m_c$ value  is  perfectly compatible with, but a factor of three  less precise than,  the 
recent result from global fits 
$m_c = 1.080(70)~{\rm GeV}$ \cite{babar2009}.
Nevertheless, the
${\cal R}$ ratio seems to be an interesting  observable since it is
primarily sensitive to phase-space ratios and is almost independent of
both perturbative and non-perturbative effects. The reduction of the
experimental error in the ${\cal R}$ ratio by a factor of three will
lead to the determination of the charm quark mass with the precision
comparable  to  the precision currently achieved in global fits.   As the final 
remark, we point out that for central values of bottom and charm quark 
masses determined from semileptonic fits \cite{babar2009}, $m_b = 4.55~{\rm GeV}$ and 
$m_c = 1.077~{\rm GeV}$, the ${\cal R}$ ratio is $0.232$, in perfect 
agreement with the ALEPH  result Eq.(\ref{eq543}). 

\section{Conclusions}
\label{sect:4}

In this paper, we studied the NNLO QCD corrections to semileptonic
$b \to X_c l \bar \nu_l$ decays. We described the computational 
method  that allows
us to consider decays into both massless and massive leptons and
impose arbitrary cuts on the final state particles.

We showed that  non-BLM NNLO  QCD corrections to $b \to X_c l \bar \nu_l$
decays, with 
$l = e,\mu$,  
are not very sensitive to cuts on the lepton energy, 
as long as the cut is below  $1~{\rm GeV}$. 
For higher values of the lepton energy cut, the non-BLM corrections do develop 
$E_{\rm cut}$ dependence, although it is not very strong.   We also found that 
there are very efficient cancellations of QCD radiative 
corrections to {\it normalized} moments, 
that are used in global fits to semileptonic  $B$-decays,  and 
that such cancellations are  crucial for making the claimed 
accuracy of the fits 
credible. We argued that there are good reasons to believe 
that such cancellations are not 
accidental, and that they persist in higher orders of perturbation 
theory as well.

We also computed  QCD radiative corrections to
the ratio of branching fractions of 
$b \to X_c \tau \bar \nu_\tau$ and 
$b \to X_c e \bar \nu_e$ decays. It turns out that radiative 
corrections to this ratio 
are very small and convergence of the perturbative expansion 
is excellent.  Since non-perturbative effects are also moderate, this ratio is,
potentially, a good source of information about bottom and charm 
quark masses.
We showed  that if the charm quark mass is extracted directly from this
ratio, the result is in good agreement  with 
the value of the 
charm quark mass obtained from fits to semileptonic $B$-decays.

\acknowledgments  We would like to thank P.~Gambino for useful 
conversations.  Our explanation of  smallness of
radiative corrections to normalized moments 
was strongly 
influenced by discussions with N.~Uraltsev. We are indebted to him 
for these and other  comments.
This research  is supported by the NSF under grant
PHY-0855365 and by the start up
funds provided by Johns Hopkins University.
Calculations reported in this paper were performed on the Homewood
High Performance Cluster of Johns Hopkins University.

\end{document}